\documentclass[sn-mathphys-num]{sn-jnl}
\usepackage{lmodern}
\usepackage{graphicx}%
\usepackage{multirow}%
\usepackage{amsmath,amssymb,amsfonts}%
\usepackage{amsthm}%
\usepackage[title]{appendix}%
\usepackage[table,xcdraw,HTML]{xcolor}
\usepackage{textcomp}%
\usepackage{manyfoot}%
\usepackage{booktabs}%
\usepackage{algorithm}%
\usepackage{algorithmicx}%
\usepackage{algpseudocode}%
\usepackage{listings}%
\usepackage{array}
\usepackage{colortbl}
\usepackage{longtable}
\usepackage{hyperref}
\usepackage{url}
\usepackage{placeins}
\newcommand{\toolname}[0]{}
\renewcommand{\toolname}[0]{{\textit{AdaptForge}}}

\begin{document}

\title[Article Title]{Adaptive and Accessible User Interfaces for Seniors Through Model-Driven Engineering}


\author*[1]{\fnm{Shavindra} \sur{Wickramathilaka}}\email{shavindra.wickramathilaka@monash.edu}

\author[1]{\fnm{John} \sur{Grundy}}\email{john.grundy@monash.edu}

\author[1]{\fnm{Kashumi} \sur{Madampe}}\email{kashumi.Madampe@monash.edu}

\author[1]{\fnm{Omar} \sur{Haggag}}\email{omar.haggag@monash.edu}

\affil*[1]{\orgdiv{Department of Software Systems and Cybersecurity, Faculty of Information Technology}, \orgname{Monash University}, \orgaddress{\street{Wellington Road}, \city{Clayton}, \postcode{3800}, \state{Victoria}, \country{Australia}}}


\abstract{The use of diverse mobile applications among senior users is becoming increasingly widespread. \textcolor{black}{However, many of these apps contain accessibility problems that result in negative user experiences for seniors.} A key reason is that software practitioners often lack the time or resources to address the broad spectrum of age-related accessibility and personalisation needs. \textcolor{black}{As current developer tools and practices encourage one-size-fits-all interfaces with limited potential to address the diversity of senior needs, there is a growing demand for approaches that support the systematic creation of adaptive, accessible app experiences.} To this end, \textcolor{black}{we present \toolname, a novel model-driven engineering (MDE) approach that enables advanced design-time adaptations of mobile application interfaces and behaviours tailored to the accessibility needs of senior users.} \toolname~uses two domain-specific languages (DSLs) to address age-related accessibility needs. The first model defines users' context-of-use parameters, while the second defines conditional accessibility scenarios and corresponding UI adaptation rules. These rules are interpreted by an MDE workflow to transform an app's original source code into personalised instances. We also report evaluations with professional software developers and senior end-users, demonstrating the feasibility and practical utility of \toolname.}

\keywords{Model-driven engineering, Adaptive user interfaces, Senior end users, Domain-specific languages, Software accessibility}



\maketitle

\section{Introduction} \label{introduction}

% According to the World Health Organisation (WHO), one in six people will be 60 years or older by 2030 \cite{who}. Many of these \textit{senior} individuals are likely to experience age-related declines in capabilities such as vision, hearing, cognition, coordination, and mobility \cite{bossini2014}. Despite being a significant demographic segment, seniors often encounter accessibility barriers in software applications, particularly in software User Interfaces (UIs) \cite{bossini2014, Paez2019, connor2017}. Our user studies with senior participants corroborate this phenomenon. For example, the persona in Figure~\ref{persona example}, based on our user studies, shows Judy, a senior with vision impairment who finds it difficult to read small text in mobile apps. To manage, she increases the text size using her device settings. However, this leads to new problems, such as misaligned UI elements and less visible information on the screen, which creates further accessibility barriers. This situation is common among older users and represents just one of many age-related needs that can negatively affect their app experience~\cite{wickramathilaka2025guidelines}.

\begin{figure}[H]
\centering
\includegraphics[width=1\textwidth]{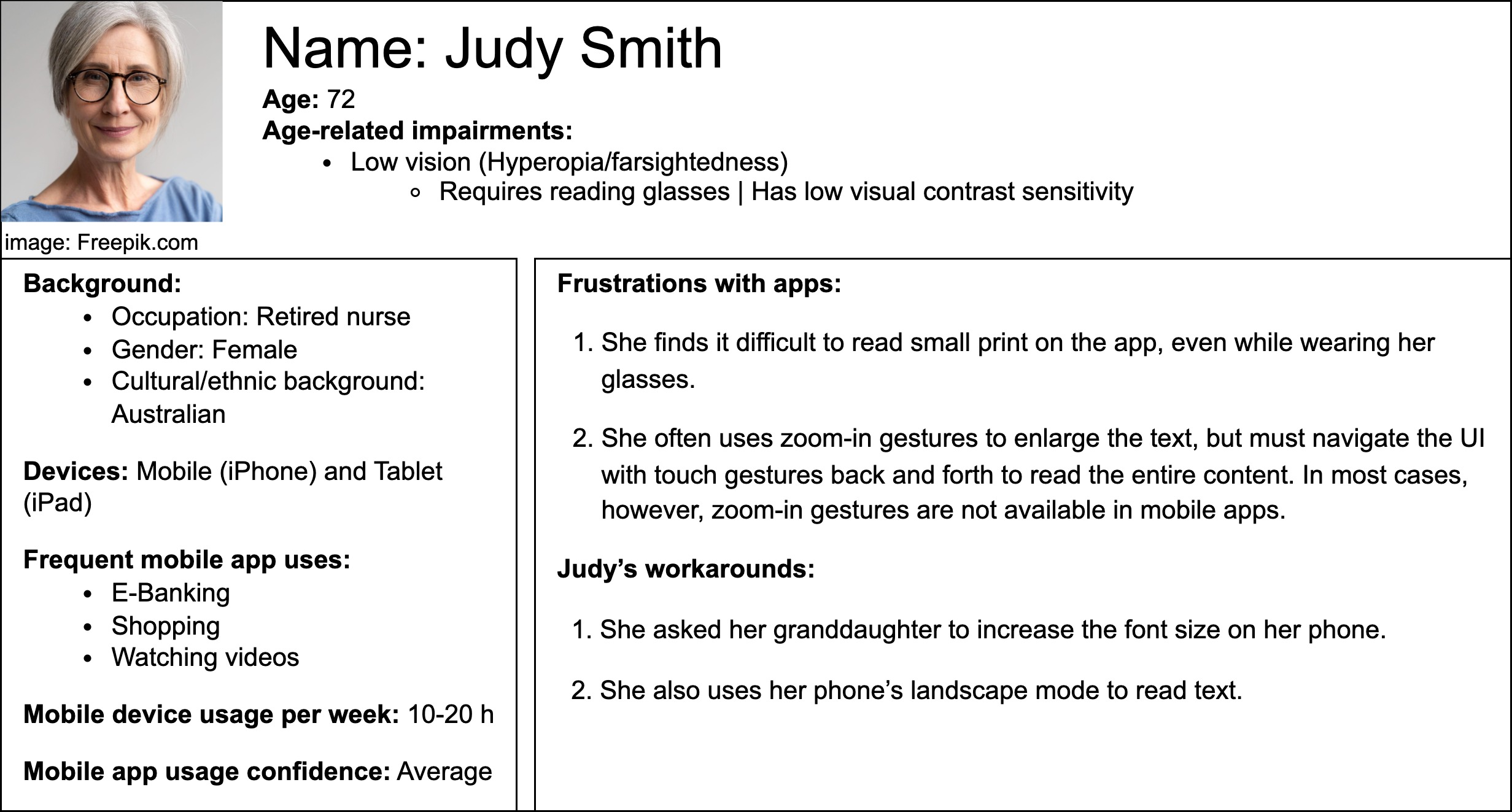}
\caption{\textcolor{black}{Meet Judy, a retired nurse who encounters significant accessibility barriers when using mobile applications due to her low vision. She struggles with small font sizes and low-contrast foreground-background colour combinations. Although she has increased the text size through her phone's global settings, this often results in misaligned UI elements and excessive vertical scrolling.}}\label{persona example}
\end{figure}

\textcolor{black}{According to the World Health Organisation (WHO), one in six people worldwide will be aged 60 or older by 2030~\cite{who}. Many of these \textit{senior} individuals are likely to face age-related challenges in vision, hearing, cognition, coordination, and mobility~\cite{bossini2014}. Despite forming a large part of the population, seniors often face accessibility issues in software applications, especially in user interfaces (UIs)~\cite{bossini2014, Paez2019, connor2017}. Our user studies support this trend. For example, the persona in Figure~\ref{persona example}, based on our research, features Judy, a senior with vision impairment who struggles to read small text in mobile apps. To cope, she increases the text size using her device settings. However, this leads to other problems, such as misaligned UI elements and less content on the screen, which create new accessibility challenges. Such situations are common among older users and reflect just one of many age-related needs that can negatively affect their experience~\cite{wickramathilaka2025guidelines}.}

A major responsibility for this issue can be attributed to software practitioners, as they often do not consider seniors as an active user community \cite{johnson2017designing}. Consequently, different age-related accessibility needs of seniors are frequently overlooked during the software development lifecycle \cite{bossini2014, Paez2019, shamsujjoha2024developer}. However, this does not imply that developers are apathetic towards the limitations faced by seniors in society. For instance, during our interview study with software developers, the majority of participants demonstrated some awareness and personal empathy for this societal challenge. For example, one participant shared their personal experience with ageing: \textit{“I'm already getting to the age where I have to take my glasses off to read up close. So, I understand some of it.”}

\textcolor{black}{Despite this awareness, applying accessibility principles in real-world software projects remains challenging. Seniors have diverse needs, and ageing is a highly individualised process, leading to significant differences in abilities, limitations, and experiences among users~\cite{czaja2007}. This variation means developers must consider numerous potential accessibility scenarios. Addressing these manually with current development tools is often time-consuming and costly, as noted by Akiki et al.~\cite{akiki2014}. One developer in our study reflected this challenge: \textit{"In day-to-day life, we tackle more with business requirements rather than user requirements [...]. Because our products change every day, our requirements change every day, and it's hard to always include all these requirements."} Therefore, any solution to this problem must also take into account the constraints faced by developers.}

A more convenient way to address accessibility challenges is by following established guidelines such as the Web Content Accessibility Guidelines (WCAG)~\cite{wcag2.2} or ISO 9241-171~\cite{iso9241}. These standards help developers design universally accessible user interfaces that aim to reduce barriers for all users, including seniors. However, this one-size-fits-all approach often falls short because seniors have diverse and non-homogeneous accessibility needs that cannot always be met through general design principles alone~\cite{akiki2014}.

A more pragmatic approach to addressing this issue is the combination of: (1) Model-Driven Engineering (MDE) \cite{brambilla2017}, which considers software models as primary artefacts for generating application code with minimal to no handwritten coding \cite{brambilla2017}; and (2) Domain-Specific Languages (DSLs) \cite{brambilla2017}, which enable developers to model the specific requirements of a given application domain -- in this case, the age-related accessibility needs of seniors. \textcolor{black}{A DSL, or a combination of DSLs, can be designed specifically for software practitioners as end-users. When the purpose of these DSLs is to capture a particular domain: such as age-related accessibility and personalisation needs, they are often more suitable than general-purpose modelling tools such as Unified Modelling Language (UML)~\cite{brambilla2017}. However, it is important to acknowledge that developers must still invest time in learning how to use such tools, despite already working under tight time and resource constraints.}

\textcolor{black}{This is where the inherent model transformation and code generation capabilities of a model-driven engineering (MDE) workflow become valuable. By integrating DSLs into an MDE process, developers can transform their models into executable applications that are both more accessible and personalised for their target users, while significantly reducing the manual effort required through automation. The effectiveness of this combined approach has been well demonstrated in several studies \cite{bendaly2018, ghaibi2017, akiki2016, yigitbas2020}.} However, these studies have not specifically addressed the needs of the senior community. As a result, several gaps remain in terms of the comprehensiveness of senior accessibility needs modelling, as well as in the demonstration and evaluation of the approach \cite{wickramathilaka2023}. With these gaps in mind, we devised the following approach to conduct our study.

\begin{enumerate}
    \item We first ran an exploratory focus group study with a community of seniors to understand their accessibility and personalisation challenges with mobile apps;
    \item \textcolor{black}{We developed a set of DSLs aimed at \textbf{software developers}, allowing them to represent senior accessibility and personalisation needs as software modelling artefacts;}
    \item We prototyped an MDE-based Flutter application automatic adaptation tool that takes our DSLs defined by developers and creates a new version of the Flutter app applying the desired UI accessibility adaptations;
    \item We evaluated our generated adapted Flutter apps with the same senior community to see how well we had addressed their accessibility needs; and
    \item We evaluated the DSLs and the overall MDE prototype with Flutter developers to see how they perceive the practicality and usefulness of such an approach.
\end{enumerate}

\noindent \textbf{\textcolor{black}{Key contributions to the MDE and Adaptive UI sub-domain, and the Low-code development community:}}
\begin{itemize}
    \item \textcolor{black}{We propose a pair of domain-specific languages (DSLs) that iteratively advance the current state-of-the-art tools for modelling age-related accessibility and adaptation needs.}
    \item \textcolor{black}{We propose a novel model-driven engineering (MDE) approach that shows how low-code tools can be effectively integrated with traditional development practices. This integration enables developers to create accessible and adaptive applications for seniors, along with other user groups, without breaching typical time or budget constraints in software projects.}
\end{itemize}

\noindent \textbf{\textcolor{black}{Key contributions to the broader Software engineering community:}}
\begin{itemize}
    \item \textcolor{black}{We demonstrate that an MDE approach such as \toolname~is a practical and useful tool for the software industry, enabling developers to create better, more accessible, and more personalised applications for disadvantaged user groups, such as seniors.}
\end{itemize}

% The key contributions that we discuss within this paper are: 
% \begin{enumerate}
%     \item How our DSLs significantly enhance the current capabilities and depth in modelling age-related accessibility and adaptation needs of seniors;
%     \item How our novel MDE-based approach enhances traditional development tools and practices by enabling developers to deliver more accessible and personalised apps for seniors while seamlessly integrating with their existing workflows; and
%     \item How we generated insights into the expectations of both developers and seniors regarding adaptive UI solutions, specifically developer perspectives, which have not been explored in other related works within this subdomain.
% \end{enumerate}

In this paper, we begin by detailing our proposed approach in Section~\ref{methodology}. We then describe two user studies conducted with senior end-users and software developers to evaluate the approach, followed by a discussion of the findings in Section~\ref{evaluation}. Section~\ref{limitations} outlines the limitations and directions for future work, and Section~\ref{related work} reviews related literature. Finally, we discuss the research contributions in Section~\ref{contributions} and conclude the paper in Section~\ref{conclusion}.

\section{Our Approach} \label{methodology}

\subsection{Research methods} \label{methods}

\begin{figure}
\centering
\includegraphics[width=1\textwidth]{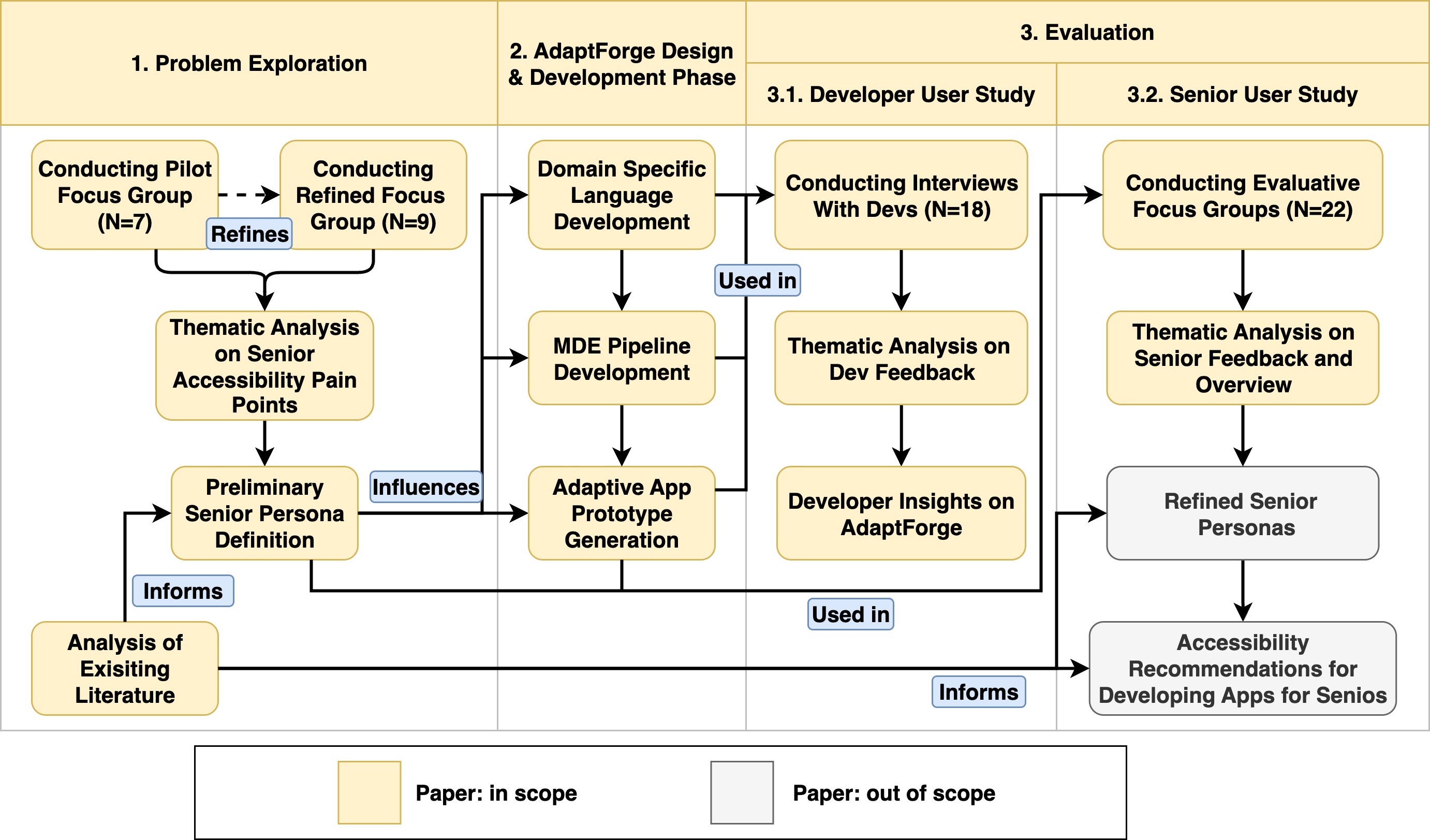}
\caption[The overall research methodology]{The overall Research Methodology}
\label{methodology_overview}
\end{figure}

This project primarily consists of 3 phases. The initial problem exploration, the design and development tasks that were informed by the exploratory insights and finally an evaluation with two different types of stakeholders to determine how successful we were in addressing our goals. The Figure \ref{methodology_overview} further illustrates individual tasks within each phase, followed by a brief phase-by-phase overview.

\subsubsection{Phase 1: Problem exploration} \label{Problem exploration}

At the outset of our study, we aimed to gain a deeper understanding of the real-world UI challenges faced by senior app users. \textcolor{black}{To this end, we conducted an initial pilot study with seven senior participants, followed by a focus group involving nine additional seniors. Overall, the mean age among these 16 participants was 73.9 years, while the median was 74. All participants were recruited from a well-established senior community in Australia named the University of the Third Age (U3A). Using thematic analysis on the qualitative data collected, we identified three key types of app adaptations required to holistically address common accessibility barriers encountered by seniors: (1) Presentation adaptations, (2) Multi-modality adaptations, and (3) Navigational adaptations.}

\textcolor{black}{To contextualise these adaptation types, we derived three preliminary personas to guide and motivate the subsequent tool design. The first persona focused on \textbf{vision-related limitations}, such as reduced contrast sensitivity, which often necessitate presentation adaptations (see Figure~\ref{persona example}). The second persona illustrated the impact of \textbf{mobility impairments}, such as arthritis, which may require alternative input and output modalities, including voice commands and screen readers. The third persona addressed \textbf{cognitive limitations}, such as memory decline, which create a need for simplified app navigation to reduce cognitive load. A detailed account of this exploratory phase and the development of these personas is provided in Wickramathilaka et al.~\cite{wickramathilaka2025guidelines}.}

\textcolor{black}{Initially, these personas served as an internal design tool to guide the development of \toolname\ and to identify complex code-generation challenges associated with addressing age-related accessibility needs. For example, insights from our exploratory focus groups revealed that increasing text size and adding new interface elements, such as buttons for voice input, could result in a more cluttered user interface with extended vertical scroll length. While such changes improve text readability, they may also introduce new barriers for seniors with cognitive limitations or reduced hand dexterity. In this context, the personas helped us identify the need for non-standard adaptations, such as implementing a wizard-style interface to segment content.}

\textcolor{black}{In a later stage, we refined and extended these personas to more comprehensively represent the accessibility and personalisation challenges faced by senior users of mobile applications, as detailed in Wickramathilaka et al.~\cite{wickramathilaka2025guidelines}. It is important to note that while the personas effectively captured the needs of a group of older users, we designed all our tools with extensibility in mind. The design process was influenced by a variety of resources such as accessibility standards (e.g., WCAG and ISO 9241-171), previous studies on app accessibility for older adults (e.g., \cite{ahmad2020, harte2017, morey2019, watkins2014}) and existing DSL metamodels for accessible and adaptive UI design (e.g., \cite{yigitbas2020, bendaly2018}). As a result, the final prototype is capable of modelling accessibility and adaptation requirements for a broader range of users beyond those represented in the three preliminary personas.}

% The findings from these studies revealed that seniors face numerous challenges when interacting with modern app interfaces. Their feedback played a pivotal role in the development of several artefacts, including personas, user narratives, and UI adaptation scenarios, which subsequently informed the design and development phases of our project. A simplified example of these personas is illustrated in Figure \ref{persona example}, providing further context for our methodology.

\subsubsection{Phase 2.1: DSL design and development} \label{DSL development}

Subsequently, we utilised the personas developed during our exploratory phase, alongside the Web Content Accessibility Guidelines (WCAG), to evaluate the current state-of-the-art approaches in the domain of Model-Driven Engineering (MDE)-based UI adaptation. The personas served as a means to assess whether existing approaches could accommodate the specific needs identified in our focus group study. Meanwhile, WCAG served as a structured baseline due to its comprehensive accessibility criteria, helping us assess how well existing approaches could be extended to meet diverse age-related accessibility and personalisation needs. Our analysis revealed that although current MDE-based approaches exhibit potential, they fall short in comprehensively modelling the diverse accessibility and personalisation needs of seniors~\cite{wickramathilaka2023}.

\textcolor{black}{However, our investigation also revealed that an effective approach to addressing accessibility and personalisation in user interfaces has already been explored in prior research. Notably, Yigitbas et al.~\cite{yigitbas2020} proposed a method that combines two domain-specific languages (DSLs) with a user interface modelling language, IFML~\cite{brambilla2014}, to capture diverse accessibility scenarios for general users and employ model-driven engineering (MDE) to transform these models into adaptive applications. We recognised the potential of this approach for our own objectives and subsequently extended it. Drawing on insights from existing literature, we developed our own interpretation of using multiple DSLs in tandem, one to abstractly represent accessibility scenarios encountered by senior users, and the other to define concrete adaptation logic. These models are then interpreted within a model-driven engineering workflow to transform application source code in a way that reflects the accessibility and personalisation needs captured in the DSLs.}

\textcolor{black}{Another reason we adopted a dual-DSL approach is its logical alignment with the goals of adaptive application design. As software practitioners, we recognised that defining adaptation rules as complex conditional statements using relational and logical operators, offers a powerful and intuitive way to represent real-world accessibility scenarios encountered by senior users. For instance, consider the persona in Figure~\ref{persona example}. We could express a condition where adaptations are triggered if the app detects that the user is over 72 years old \textit{and} has a low vision impairment. Based on this condition, we could specify the appropriate adaptation logic required to generate a more tailored application instance for that user. This is precisely the purpose of one of our domain-specific languages: \textit{Adapt DSL}.}

\textcolor{black}{However, such conditional expressions require access to contextual data to function effectively. A rule on its own cannot determine whether a user is over 72 or what impairments they may have. To address this, we introduce a second DSL: \textit{Context DSL}, which models the relevant accessibility-related parameters of the user. In designing Context DSL, we recognised that simply recording a few user details such as age and device type is insufficient. Our goal was to ensure extensibility across all senior demographics and application domains (e.g., e-health, e-banking, retail). Since the intended users of these tools are software practitioners working in diverse domains, we prioritised flexibility and extensibility. As such, Context DSL supports the modelling of a comprehensive range of parameters across user, platform, and environmental contexts to accommodate a wide array of accessibility and personalisation scenarios in tandem with Adapt DSL.}

% As a potential solution, we designed and developed two novel Domain-Specific Languages (DSLs) -- namely, Context DSL and Adapt DSL -- to effectively capture age-related accessibility and adaptation requirements during the application design phase. Our Context DSL is designed to capture the accessibility parameters of senior users, following a semi-graphical tree structure that allows developers to comprehensively model and create personas of senior users, either individually or by segment. In contrast, our Adapt DSL serves as a companion textual modelling language that references the information stored in the Context DSL model. It functions as a rules engine to conditionally generate or modify an application's source code, ensuring that the UI is tailored to the specific accessibility needs of senior users.

\subsubsection{Phase 2.2: MDE based Flutter prototype development} \label{prototype development}

% To realise our novel DSLs and our overall approach, we developed a proof-of-concept tool suite, \toolname, using the Eclipse Modelling Framework \cite{emf} and the Flutter Framework \cite{flutter}. The decision to use Flutter is influenced by 2 factors: 1) We wanted to develop our adaptive app prototypes using a widely used cross-platform app development framework. Doing so would enable us to understand how to apply our model-driven engineering to one device type (mobile) in the beginning and then generalise our lessons learned to other devices and platforms as well. Furthermore, it would allow us to tap into a global developer community with a wide range of experiences in software development during the evalaution stage. 2) The choice of Flutter over another similar framework such as React Native was arbitrary as authors already had prior experience with making Flutter apps adaptive.

% Therefore, our Model-Driven Engineering pipeline would take an Adaptation Rules model defined using Adapt DSL and a Context-of-Use model defined in Context DSL that represents the accessibility and adaptation needs of a single or a group of users along with a Flutter app defined in the Dart programming language. Based on the adaptation logic defined in our adaptation rules, the Flutter app's source code is then altered whenever the conditional statement that represents the accessibility scenario of the end user is triggered. This process is further illustrated in Figure \ref{overall_architecture}.

To realise our novel DSLs and overall approach, we developed a proof-of-concept tool suite, \toolname, using the Eclipse Modelling Framework (EMF)~\cite{emf} and the Flutter framework~\cite{flutter}. \textcolor{black}{The decision to adopt Flutter was influenced by two key factors: (1) we aimed to develop adaptive app prototypes using a widely adopted cross-platform framework. This choice allowed us to initially apply our model-driven engineering (MDE) approach to mobile applications, to generalise our findings to other platforms (e.g., desktop and web-based) in the future. Additionally, it facilitated access to a large global developer community with diverse experiences, which proved valuable during the evaluation stage. (2) Our selection of Flutter over other frameworks, such as React Native, was primarily due to the authors' prior experience with implementing adaptive behaviours in Flutter applications.}

Our MDE pipeline integrates two DSL inputs: an Adaptation Rules model defined using Adapt DSL and a Context-of-Use model defined using Context DSL. These models collectively represent the accessibility and personalisation needs of an individual or a group of users. The pipeline also requires a Flutter application written in the Dart programming language. When the conditional statements in the adaptation rules match specific context scenarios, the Flutter app's source code is modified accordingly. This transformation process is illustrated in Figure~\ref{overall_architecture}.

% This suite enables Dart-based source code adaptations for any app developed using the Flutter framework. Developers can define a diverse range of age-related needs of their senior end-users through the two DSLs at the app design stage, which then serve as input to tailor the source code according to the specified models. 

\subsubsection{Phase 3: Evaluation} \label{}

In the final stage of our study, we conducted two separate user studies to evaluate our tool suite: (1) a developer interview study, involving 18 developers with professional experience with the Flutter framework; and (2) a focus group study/acceptance test, conducted with 22 participants across three focus groups from the same senior user community we collaborated with during the exploratory stage. The objective of the senior user study was to assess whether the generated adapted UI instances effectively address the accessibility barriers faced by seniors due to their age-related needs. In this paper, we place greater emphasis on the developer study evaluation, as it provides critical insights into the effectiveness of our novel approach and highlights areas for improvement. This focus is particularly important from the perspective of our primary audience -- the software developer community.

\subsection{Overview of ~\toolname} \label{overview}

\subsubsection{\toolname~architectural components}

\begin{figure}
\centering
\includegraphics[width=1\textwidth]{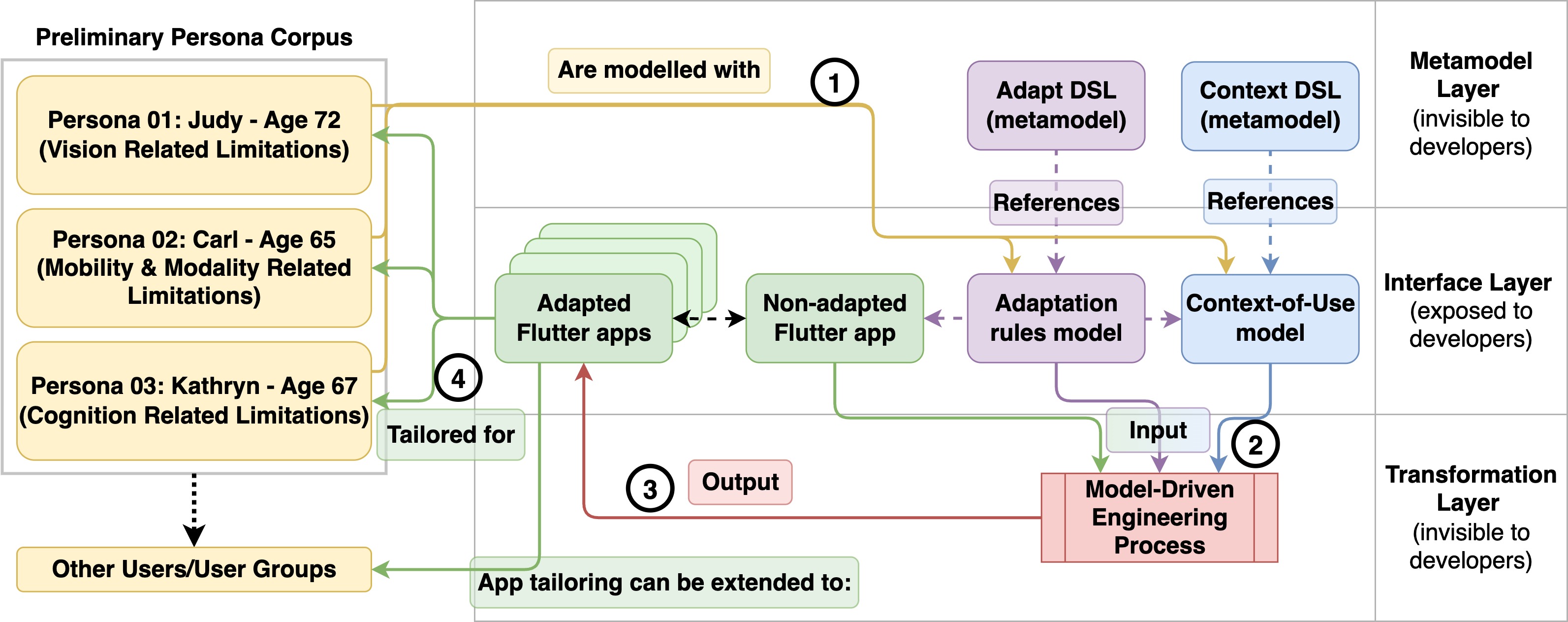}
\caption[The overall architecture of \toolname]{The overall architecture of \toolname. \textcolor{black}{Note that the scalability of tailored app instance generation is not restricted to the three personas in the preliminary persona corpus; these are illustrative examples of how the overall approach can tailor personalised app instances for a variety of senior needs. The modelling of accessibility needs and app personalisations can be extended to other users and user groups (senior or otherwise).}}
\label{overall_architecture}
\end{figure}

% \footnotetext{The images of seniors used in the figure are taken from Freepik.com.}

In our proposed \toolname~approach (illustrated in Figure \ref{overall_architecture}), there are three architectural layers: (1) the Metamodel Layer, (2) the Interface Layer, and (3) the Transformation Layer. 

\begin{enumerate}
    \item \textbf{Metamodel Layer:}  
    This layer comprises two \textit{Domain-Specific Languages (DSLs)}: \textit{Adapt DSL} and \textit{Context DSL}, which define the grammar and syntax for their respective model instances in the Interface Layer. These DSLs serve as the foundation for capturing accessibility-related requirements and adaptation rules.
    
    \item \textbf{Interface Layer:}  
    The Interface Layer provides developers with an interactive environment to utilise the two DSL model instances: the \textit{Adaptation Rules Model} and the \textit{Context-of-Use Model}. These models enable developers to comprehensively represent various accessibility scenarios for senior users or other user segments. Additionally, developers must ensure the integration of the DSL models into the Flutter application’s source codebase, which facilitates the adaptation of UI elements based on user-specific accessibility needs.
    
    \item \textbf{Transformation Layer:}  
    Once the DSL models are integrated, developers can trigger the UI adaptation logic defined within them. The Transformation Layer is responsible for performing \textit{model-to-text transformations}, automatically generating alternative Flutter app UI versions that align with the user requirements and accessibility scenarios specified in the DSL models.
\end{enumerate}

\subsubsection{\toolname~workflow in a real-world scenario}

% Aside from the tool interface layer, the other two layers in Figure~\ref{overall_architecture} are either hidden or act as black boxes to the \textbf{users of \toolname: software developers}. To explain the workflow more clearly from a developer's point of view, let us consider an example. Suppose a development team has built an e-banking Flutter application for a user base that includes many senior users. They could first either map context-of-use models for each user or opt for a more segmented approach where they create and model a few personas to holistically represent the user base (in Figure \ref{overall_architecture} example, we use the preliminary persona corpus derived during the problem exploration stage: Judy -- age 72, Carl -- age 65, and Kathryn -- age 67 as abstract examples) using Context DSL. Then they can integrate their e-banking app into the \toolname's MDE pipeline and begin to express adaptation rules to be applied on certain widget within the app, when an adaptation condition is triggered (i.e., IF the user is aged above 60 AND has a low visual contrast sensitivity condition, switch the app theme to black and white and increase the font-weight attribute of all the text widgets in the app 'bold'). 

\textcolor{black}{Aside from the tool interface layer, the other two layers in Figure~\ref{overall_architecture} are either hidden from or operate as black boxes to the \textbf{users of \toolname: software developers}. To clarify the workflow from a developer’s perspective, consider the following example.}

\textcolor{black}{Suppose a development team has built an e-banking application using Flutter for a user base that includes many senior users. To begin incorporating accessibility-adaptations needs, the team can first model the users' context-of-use parameters using \toolname’s Context DSL. This can be done either by mapping detailed context-of-use models for individual users or by taking a more segmented approach of creating a small number of representative personas. In the example shown in Figure~\ref{overall_architecture}, we use personas derived from our preliminary research phase: Judy (age 72), Carl (age 65), and Kathryn (age 67) to demonstrate the latter approach.}

\textcolor{black}{Once the context-of-use parameters of the user base are modelled with Context DSL, the team integrates their e-banking app into \toolname’s MDE pipeline. They can then begin expressing adaptation logic that defines how the app should behave under specific user conditions with Adaptation Rules models(with Adapt DSL editor). For instance, if we consider the persona example of Judy from Figure \ref{persona example}, developers might specify a rule such as:  \textit{IF the user is over 60 AND has low visual contrast sensitivity, THEN switch the app theme to black and white AND increase the font weight of all text widgets to 'bold'}. These rules are then automatically interpreted by the MDE workflow, enabling the generation of adapted app variants tailored to different user needs without requiring manual interface redesign.}

\textcolor{black}{The generated app variants are kept separate from the original source code of the e-banking Flutter app, which remains the primary source of truth. Upon deployment, the appropriate app variant is served to each end-user based on their accessibility limitations or personalisation preferences. It is important to note that the current prototype implementation of \toolname~does not handle the full complexity of app deployment and maintenance. However, our developer user studies have helped identify several viable strategies to support these processes, which we outline as part of our future work.}

\textcolor{black}{Ultimately, \toolname~enables software development teams to either use a newly built Flutter app or use an existing one, plug it into our MDE workflow, model user accessibility and personalisation needs using DSLs, and automatically generate adapted app variants. This process eliminates the need for teams to invest additional time and resources in manually producing multiple versions of the app. Once generation is complete, developers can deliver tailored app experiences to end-users while preserving a pristine source-of-truth codebase.}

\subsection{\toolname~Domain Specific Languages} \label{novel dsls}

\subsubsection{Context DSL} \label{context dsl}

\begin{figure}
\centering
\fbox{\includegraphics[width=0.96\textwidth]{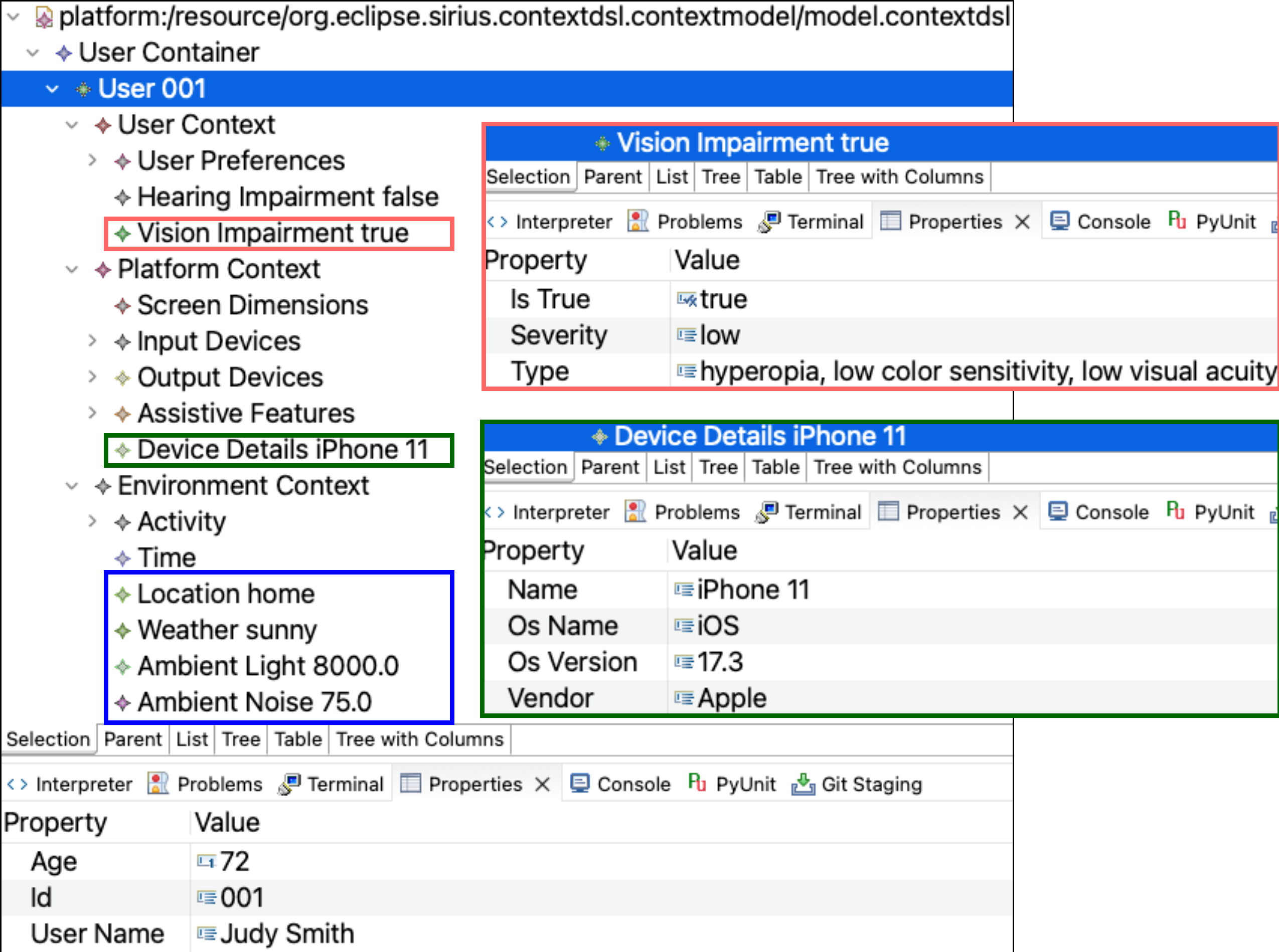}}
\caption{An example Context-of-use model for the senior persona in \toolname~for senior user Judy. Here we show a semi-graphical editor developed using Eclipse's Sirius modelling workbench.}\label{context model example}
\end{figure}

The term \textit{Context-of-use} was initially coined by Stephanidis et al. \cite{stephanidis2000} and Calvary et al. \cite{calvary2002} to explain how modern-day UI requirements are multidimensional and highly context-dependent. In particular, according to Calvary et al. \cite{calvary2002}, Context-of-use is an n-tuple that comprises the following entities: 1) \textbf{user} of the system and their attributes (e.g. accessibility needs, digital literacy, and language preferences), 2) \textbf{platform}: hardware and software components that the users use to interact with the system (e.g. device type, screen dimensions, and I/O components), and 3) \textbf{environment} in which interactions between the user and the system occur (e.g. ambient noise and lighting levels, or the weather). The concept is fairly simplistic but allows a great deal of flexibility when attempting to model an accessibility scenario of an end-user. In addition, most existing studies in the domain already champion the use of Context-of-Use modelling for adaptive UIs \cite{ghaibi2017, akiki2014, yigitbas2020, braham2022, bacha2011, bongratz2012}. 

\begin{figure}
\centering
\includegraphics[width=1\textwidth]{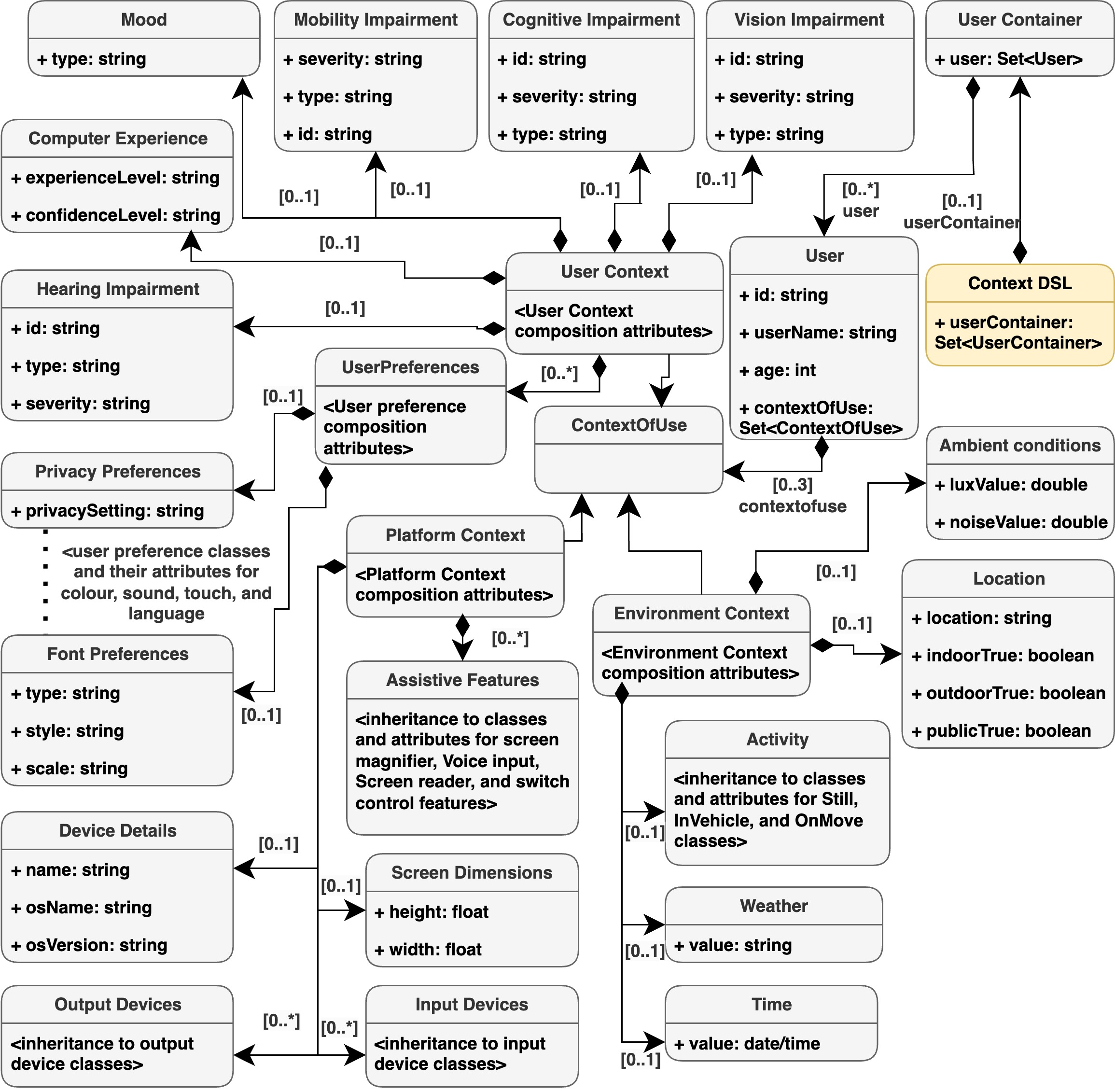}
\caption{The Context DSL metamodel. \textcolor{black}{At the root of the class diagram hierarchy is the \textit{ContextDSL} metaclass. It has a composition relationship with the \textit{UserContainer} class, which is designed to hold multiple instances of the \textit{User} class. The \textit{User} class links to an abstract \textit{ContextOfUse} class, which is further specialised into three subclasses: \textit{UserContext}, \textit{PlatformContext}, and \textit{EnvironmentContext}. In a runtime context-of-use model instance, a developer can instantiate User, Platform, and Environment context nodes under a selected user and populate them with various subnodes in a layered tree-node hierarchy, as illustrated in the metamodel.}}\label{context dsl metamodel}
\end{figure}

\textcolor{black}{To illustrate how an accessibility scenario for a senior individual can be captured, consider the persona introduced in Figure~\ref{persona example}. Judy, a 72-year-old with low colour contrast sensitivity, struggles to distinguish between foreground and background elements in mobile applications when the colour combinations are not optimised for her condition. In one scenario, Judy is using a shopping app on her iPhone 11 while sitting in her garden on a bright, sunny day. The combination of inadequate colour contrast in the app and intense ambient light makes it difficult for her to view the screen clearly. With limited options, such as shading the screen with her hand or moving to a shaded area, this situation highlights the importance of adaptive solutions.}

\textcolor{black}{This scenario can be effectively represented using a Context-of-Use model, illustrated in Figure~\ref{context model example}. The model follows a semi-graphical tree-node structure with a top-level \textit{user container} node that can represent one or more child users nodes. In this example, we define the context parameters for Judy (User ID 001). A developer can populate the model with predefined nodes such as \textit{vision impairment}, specifying attributes like severity and type (highlighted in red). Platform context, such as \textit{device details} (highlighted in green) and screen dimensions, can also be included. Environmental parameters, including \textit{location} or \textit{time}, can be populated using real-time sensor data from the user’s device (highlighted in blue).}

\textcolor{black}{Context DSL as depicted by its metamodel depicted in Figure \ref{context dsl metamodel}, is an iterative improvement upon current state-of-the-art DSL methods for modelling context-of-use scenarios. Our approach draws particular inspiration from Bendaly Hlaoui et al.~\cite{bendaly2018}, whose work presented one of the most comprehensive context-of-use DSL metamodels. We also took influence from Yigitbas et al.~\cite{yigitbas2020}; although their context-of-use metamodel was comparatively shallower, it offered valuable insights into how an additional DSL could be employed to transform static, tree-like contextual parameters into meaningful accessibility scenarios.}

\textcolor{black}{A minor enhancement in our proposed metamodel (Figure~\ref{context dsl metamodel}), compared to the related works reviewed in Wickramathilaka and Mueller~\cite{wickramathilaka2023}, is the inclusion of meta-classes for modelling user preferences within the \textit{UserContext} meta-class (e.g., preferences for privacy, colour, font, and multimedia). Additional improvements include meta-classes for assistive feature usage and I/O device modelling. Importantly, our prototype implementation is novel and distinct from existing MDE-based UI adaptation approaches. For example, Yigitbas et al.~\cite{yigitbas2020} implemented their context-of-use models using a textual, JSON-like DSL, while Bendaly Hlaoui et al.~\cite{bendaly2018} utilised a graphical DSL. In contrast, our implementation adopts a semi-graphical approach (see Figure~\ref{context model example}) built using the Sirius DSL definition workbench~\cite{sirius}. This choice reflects our design intent for the tool to be used by software developers familiar with hierarchical, semi-graphical structures (e.g., file system hierarchies). Unlike prior studies, we also evaluated the usability of this modelling approach and were able to validate our intuition through empirical feedback.}

% Our \toolname's~\textit{Context DSL} is heavily influenced by the existing studies we analysed in Wickramathilaka and Mueller \cite{wickramathilaka2023}. %The literature has extensively explored the modelling of accessibility-related scenarios. 
% By analysing these existing metamodels, we were able to design a modelling language that specifically addresses the accessibility needs of senior users. This focus is unique, as previous approaches did not consider seniors as their primary target beneficiaries. We incorporated relevant metamodel elements from existing studies into our own and made improvements, such as adding more depth to user preference modelling (e.g., privacy, colour, font, and multimedia preferences), assistive features, and I/O device modelling. The resulting DSL metamodel is illustrated in Figure \ref{context dsl metamodel}. 

\subsubsection{Adapt DSL}

While our \textit{Context DSL} can model diverse accessibility scenarios, another DSL is required by \toolname~to specify and apply adaptation operations when a context-of-use scenario is identified. For this purpose, we introduce \textit{Adapt DSL}. Unlike the abstract and graphical approach taken with Context DSL, Adapt DSL is descriptive and concrete, focusing on the definition of adaptation operations and rules. Consequently, a textual language was chosen, leveraging conditional statements commonly found in programming languages to make it intuitive for its intended users: software developers.

\begin{figure}
\centering
\includegraphics[width=0.9\textwidth]{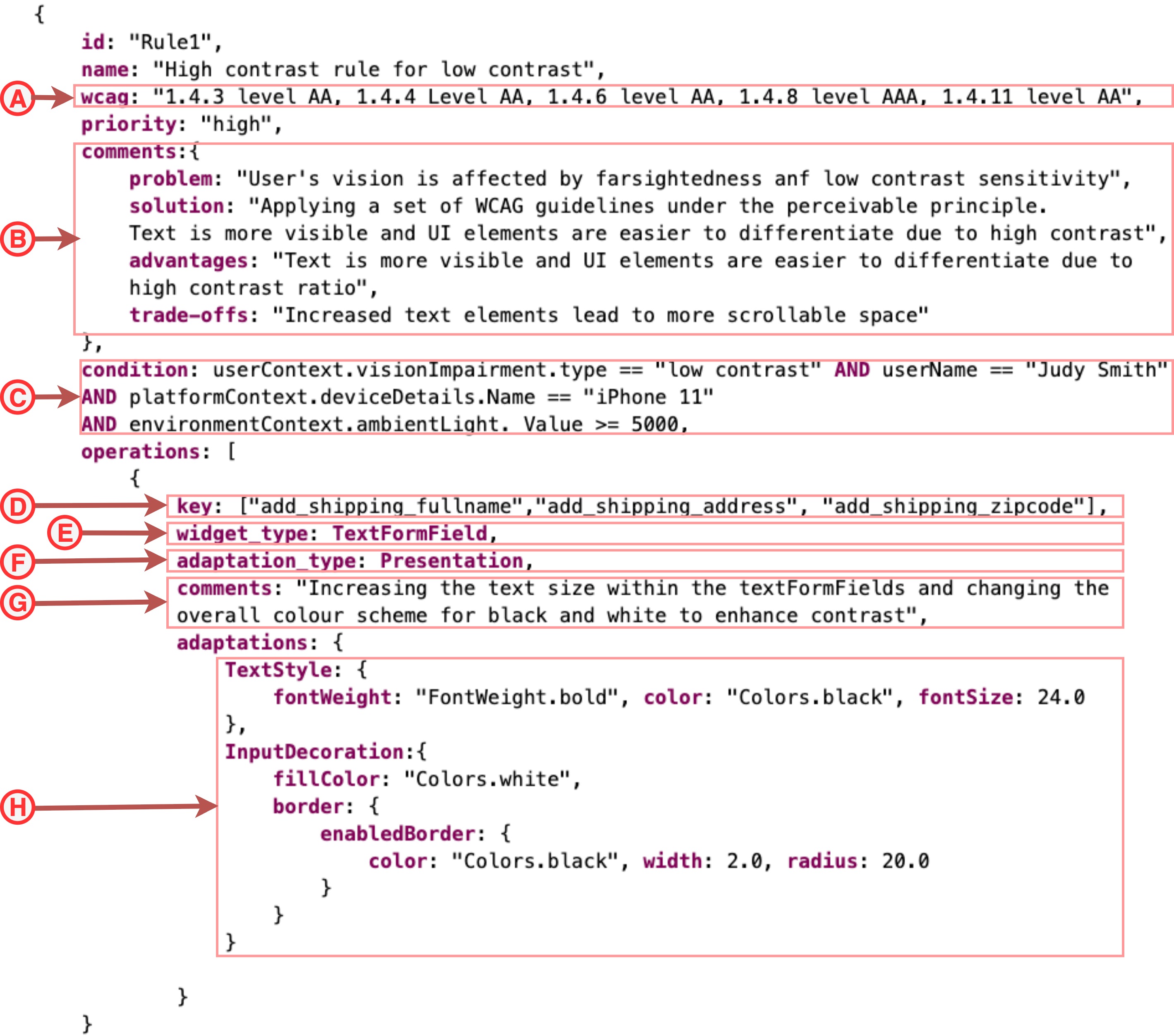}
\caption{An example adaptation rules model for Judy (a runtime model instance of Adapt DSL). To explain each section in the model: A) WCAG reference field, B) Overall comment definition of the rule, C) Context DSL referencing conditional statement, D) Unique widget(s) reference field, E) Widget type field, F) Adaptation type field, G) Comment definition field for a granular adaptation operation, and H) Example adaptation operations for a TextFormField presentation adaptation.}\label{rules model example}
\end{figure}

\textcolor{black}{
To illustrate the capabilities of \textit{AdaptDSL}, consider the user persona shown in Figure~\ref{persona example}. As discussed in the previous subsection, runtime instances of \textit{ContextDSL} can model accessibility scenarios experienced by senior end-users. These scenarios can then be referenced in \textit{AdaptDSL} through complex conditional expressions. For example, the adaptation rule illustrated in Figure~\ref{rules model example} uses the \texttt{condition} attribute (Figure~\ref{rules model example}, Section C) to incorporate various User, Platform, and Environment context parameters that define a context-of-use (e.g., a low-vision senior using an iPhone 11 in a high ambient light environment).}

\textcolor{black}{
To enhance the interpretability of each rule, developers can include metadata such as the relevant Web Content Accessibility Guidelines (WCAG) success criterion via the \texttt{wcag} attribute (Figure~\ref{rules model example}, Section A), and provide descriptive comments using the \texttt{comment} structure, which supports fields for \texttt{problem}, \texttt{solution}, \texttt{advantages}, and \texttt{trade-offs} (Figure~\ref{rules model example}, Section B).}

\textcolor{black}{The \texttt{operations} attribute then specifies the actual adaptation logic to be applied within a Flutter application. This requires several parameters: the \texttt{key} attribute (Figure~\ref{rules model example}, Section D) identifies the widget(s) to be modified; \texttt{widget\_type} (Section E) specifies the type of widget; \texttt{adaptation\_type} (Section F) defines the type of source code adaptation; and an additional \texttt{comments} attribute (Section G) allows for more granular documentation of each adaptation operation. Finally, the specific UI modifications are defined in the \texttt{adaptations} attribute (Figure~\ref{rules model example}, Section H), which supports various Flutter widgets covered by our DSL.}

\begin{figure}
\centering
\includegraphics[width=1\textwidth]{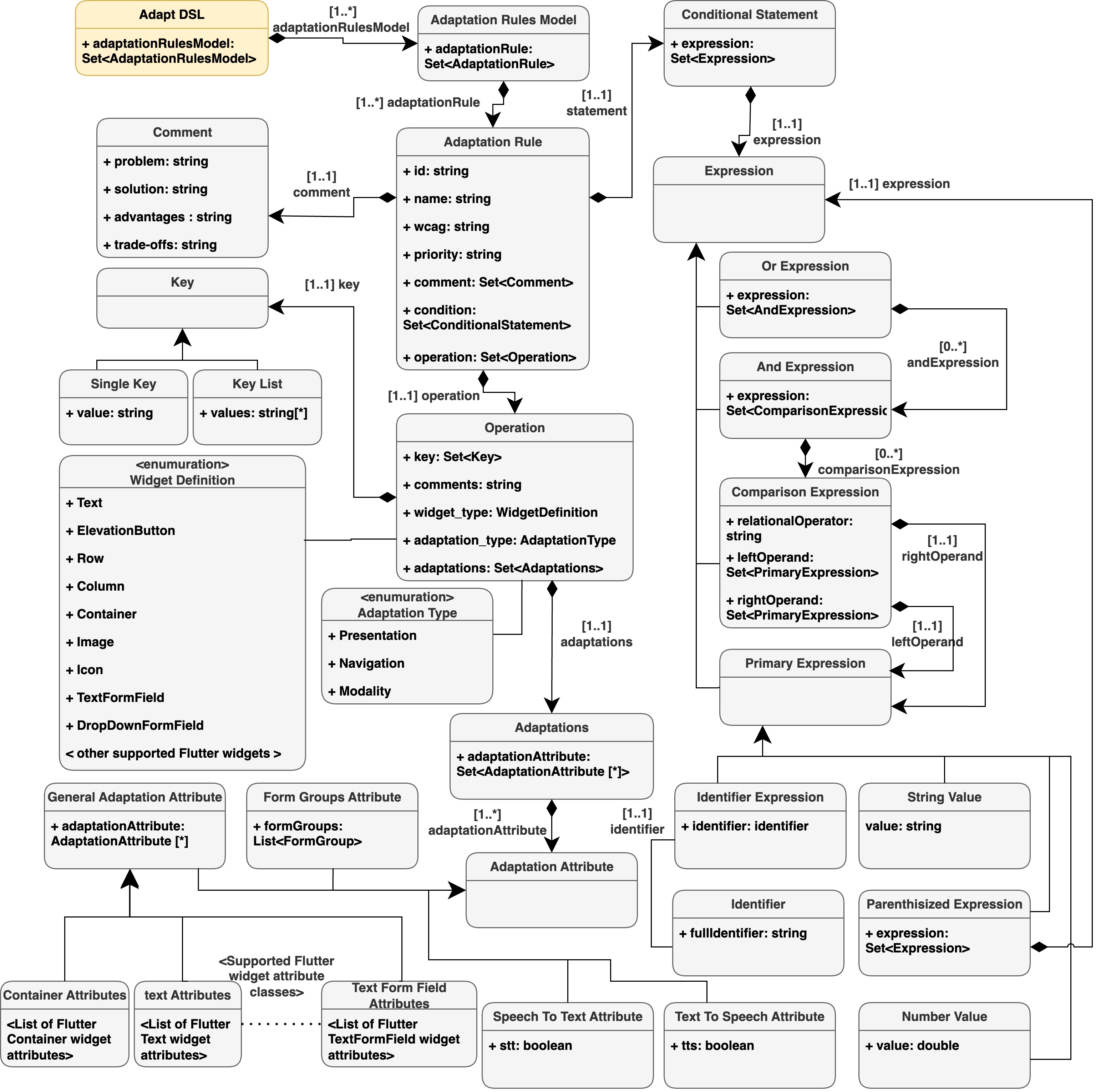}
\caption{The Adapt DSL metamodel. \textcolor{black}{At the root of the class diagram is the \textit{AdaptDSL} meta-class, which has a composition relationship with the \textit{Adaptation Rules Model} class. This allows users to instantiate multiple \textit{Adaptation Rule} objects within the Adapt DSL editor. The \textit{Adaptation Rule} class includes three key meta-classes: (1) \textit{Comment}, used to define metadata for adaptation rules; (2) \textit{Conditional Statement}, which enables users to construct complex conditional logic; and (3) \textit{Operation}, where users define concrete app adaptations targeting supported Flutter widget meta-classes such as \textit{Text}, \textit{Container}, and \textit{Icon}.}}
\label{Adapt DSL metamodel}

\end{figure}

\textcolor{black}{Our Adapt DSL metamodel is shown in Figure~\ref{Adapt DSL metamodel}, and its overall design is influenced by several existing studies~\cite{bendaly2018, yigitbas2020, bongratz2012, minon2015}. However, it also introduces several significant contributions to the research space. First, it supports the definition and referencing of various types of metadata related to specific accessibility barriers. Notably, we introduce new meta-classes that incorporate references to the Web Content Accessibility Guidelines (WCAG), enabling developers to ground their adaptation rules in widely accepted accessibility standards. Additionally, the metamodel includes a dedicated structure for documenting adaptation-related metadata such as the problem addressed, proposed solution, associated advantages, and trade-offs.}

\textcolor{black}{Another key contribution is the demonstration that abstract adaptation rule models can be integrated with widely used industrial frameworks such as Flutter, rather than relying solely on UI modelling languages like IFML to identify adaptation points \cite{yigitbas2020}. We show that any framework exposing granular UI components can potentially be integrated with Adapt DSL at the metamodel level, allowing flexible adaptation of individual components or groups of components based on developer-defined rules. Finally, we believe our prototype implementation of adaptation rules represents one of the most mature realisations of this concept among the MDE-based adaptive UI approaches we examined in our prior study~\cite{wickramathilaka2023}.}

% These include the ability to define and reference WCAG guidelines related to the adaptation rules being established, the capability to define and maintain a documentation thread for each major adaptation rule and its more granular adaptation operations, and the design of the DSL around an established professional software development tool such as Flutter to demonstrate the broader applicability of this approach. Additionally, the \toolname's Adapt DSL is more advanced than similar tools in the literature, particularly in aspects such as the flexibility and granularity of UI adaptation operations and its robustness and generalisability when applying adaptations across a wider range of apps.

\subsection{MDE-based Adaptive user interface modification}

Once the developer has modelled the age-related accessibility adaptation needs of their senior user base using \toolname's \textit{Context DSL} and \textit{Adapt DSL}, these models are transformed into structured data formats and then fed into a Model-Driven Engineering (MDE) pipeline. \toolname's MDE pipeline also takes an off-the-shelf Flutter application source code to adapt as its input.

\begin{figure}
\centering
\includegraphics[width=1\textwidth]{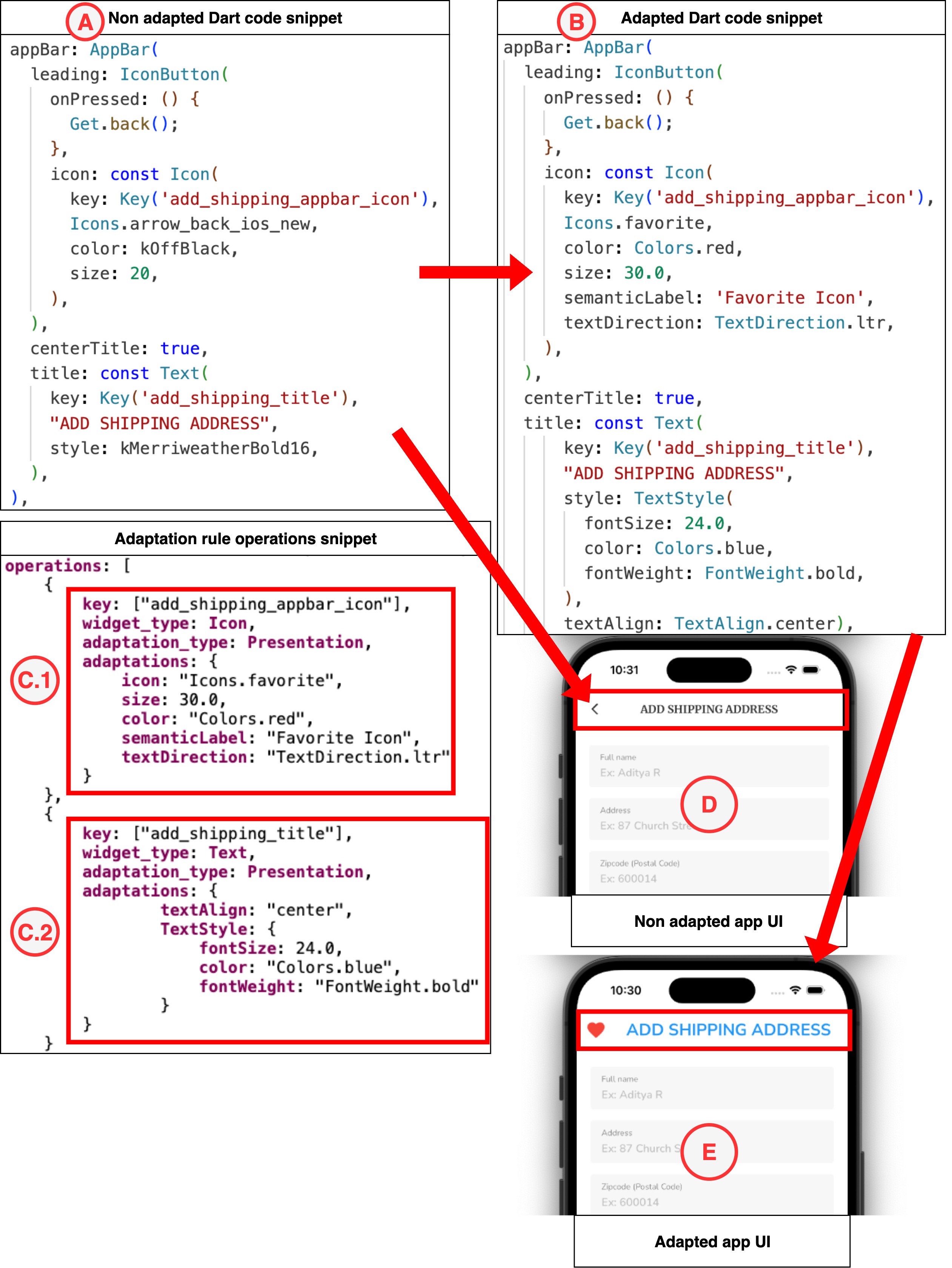}
\caption{\textcolor{black}{A simple source code transformation example of a \textbf{presentation adaptation}. \textcolor{black}{In this example, two Flutter widgets (Text and Icon) within an AppBar are transformed. The original app (D) and its source code (A) were sourced from an open-source GitHub repository. The MDE pipeline applies the specified adaptation rules operations (C.1 and C.2) to modify the property values of each Flutter widget (B). The adapted app instance (E) now reflects the changes defined in (C.1) -- Icon widget and (C.2) -- Text} widget.}}\label{pipeline example}
\end{figure}

The \toolname~MDE pipeline queries the adaptation rules model for key properties within a rule's adaptation operations. Upon identifying these key properties, it traverses the Abstract Syntax Tree (AST) of the Flutter app's Dart source code to locate the UI widgets with the specified key properties. Once these widgets are identified, the MDE pipeline applies source code modifications as defined by the adaptation rules model. An example of this adaptation pipeline is illustrated in Figure \ref{pipeline example}.

Both our \textit{Adapt DSL} and the MDE pipeline are tightly integrated with the Flutter framework in our \toolname~prototype proof-of-concept tooling. Consequently, porting our current implementation to a different tool stack (e.g., React, Angular) will require significant rework of both components to their different UI languages. However, our proposed approach is designed to be highly flexible for applications developed within the Flutter framework. Currently, most demonstrable app adaptations can be applied with minimal integration-related changes to the original codebase. For instance, all screenshots generated for this paper were taken from an open-source Flutter app project\footnote{\url{https://github.com/adeeteya/FlutterFurnitureApp}}, in which we had no involvement in development. We simply loaded the app's source code into \toolname~to demonstrate its adaptability through our approach. For example, to adapt the presentation of both the Text and Icon widgets in one of the app’s UIs (Figure \ref{pipeline example}), only two additional lines of code were required -- key property lines enabling the MDE pipeline to query the Abstract Syntax Tree (AST) and identify the relevant widgets. We believe that most apps developed using Flutter can be similarly adapted with minimal code modifications.

\subsection{An Example Usage} \label{examples}

Drawing from insights gained through our exploratory focus group studies with seniors, we identified three categories of adaptations required to address their accessibility and personalisation needs: : Presentation, Modality, and Navigation. In this section, we use this internal classification framework to provide use cases and explain how a software developer can use the \toolname~prototype to adapt application source code, resulting in tailored, adaptive user interface instances.

% As such, there are three types of app adaptations supported by the current proof-of-concept: Presentation, Modality, and Navigation adaptations. Each of the three is explained with examples below.

\subsubsection{Presentation adaptation}

\begin{figure}
\centering
\includegraphics[width=1\textwidth]{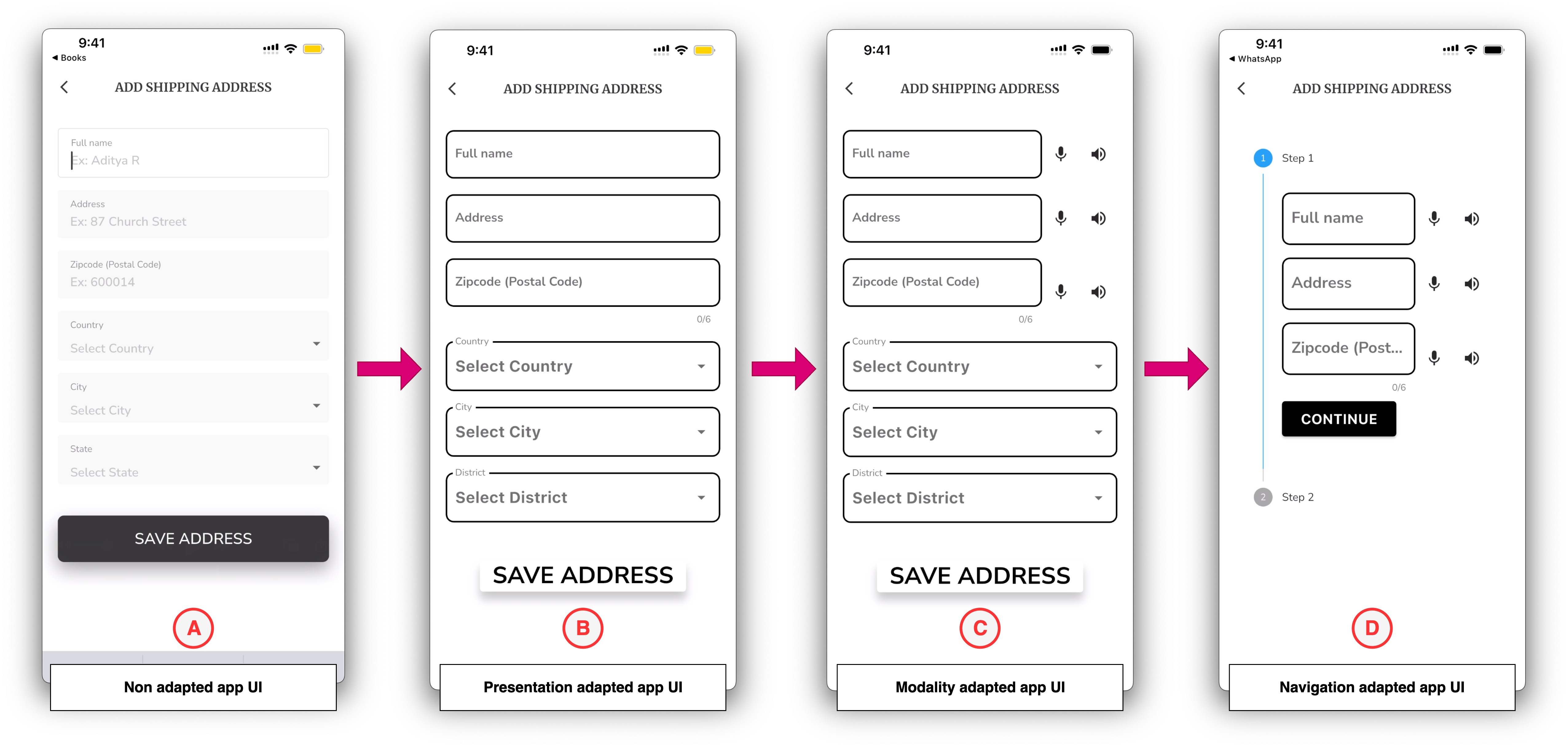}
\caption{An example is illustrated where the \textit{Add Shipping Address} interface of the open-source furniture app, shown in (A), is incrementally adapted using \toolname. First, the UI widget presentation attributes are modified, as seen in (B). Next, text-to-speech and speech-to-text functionalities are integrated, as depicted in (C). Finally, the adapted source code is further modified to transform the form into a dynamic stepper form, resulting in a navigation-adapted UI instance, shown in (D).}\label{adapted app instances example}
\end{figure}

\textcolor{black}{In the example persona of Judy given in Figure \ref{persona example}, we describe how her vision impairment creates accessibility barriers in her app UIs due to small text and insufficient background-foreground colour contrast. However, with the presentation adaptation capabilities of \toolname, we demonstrate how our approach can mitigate such issues. Ideally, to meet such a requirement, our adapted app instance should be similar to Section (B) in Figure \ref{adapted app instances example}.}

\textcolor{black}{In this section, however, the purpose is to depict how we achieve such presentation adaptations through a concrete example. To do so, let us consider the least amount of presentation adaptations that we can achieve through \toolname: a simple widget attribute change to two of the most commonly used Flutter widgets: A Text widget and an Icon widget. In Figure \ref{pipeline example}, we illustrate a non-adapted app instance in Section (D). Among the granular widgets that aggregate the UI, let us consider the Text widget that acts as the page header: 'ADD SHIPPING ADDRESS' and the 'back' icon given in the non-adapted UI. For the sake of a demonstration, let us define two adaptation operations within an adaptation rule: (C.1) aims to change the 'back' icon into a 'favourite' icon, whereas we use (C.2) to increase the text size and change the colour of the header text widget. Once these adaptation operations are executed within our MDE pipeline, Section (B) depicts how the 'add\_shipping\_appbar\_icon' and 'add\_shipping\_title' widgets now have modified attributes compared to the non-adapted UI source code from Section (A). A comparison of Section (D) and Section (E) reveals how the source code changes have changed the UI.}

\textcolor{black}{It is worth noting that we used only Text and Icon widgets in our demonstration as they can be easily and visually demonstrated. However, our current prototype supports similar widget attribute adaptations for a range of Flutter widgets, including Container, ElevatedButton, TextField, TextFormField, DropdownField, Row, Column, AppBar, Image, GridView, and SliverGrid. In the current iteration, we support more commonly used Flutter widgets, but the adaptation capabilities can be easily extended to others.}

Developers can add a \textit{key} property with a unique identifier to any of the aforementioned Flutter widgets and then use the Adapt DSL's rules model to specify stylistic property modifications for specific contextual situations, as modelled by the Context DSL. The \toolname~code generator subsequently injects these modified properties into the relevant widgets, updating the source code accordingly.

\subsubsection{Multi-modality adaptation}

\begin{figure}
\centering
\includegraphics[width=1\textwidth]{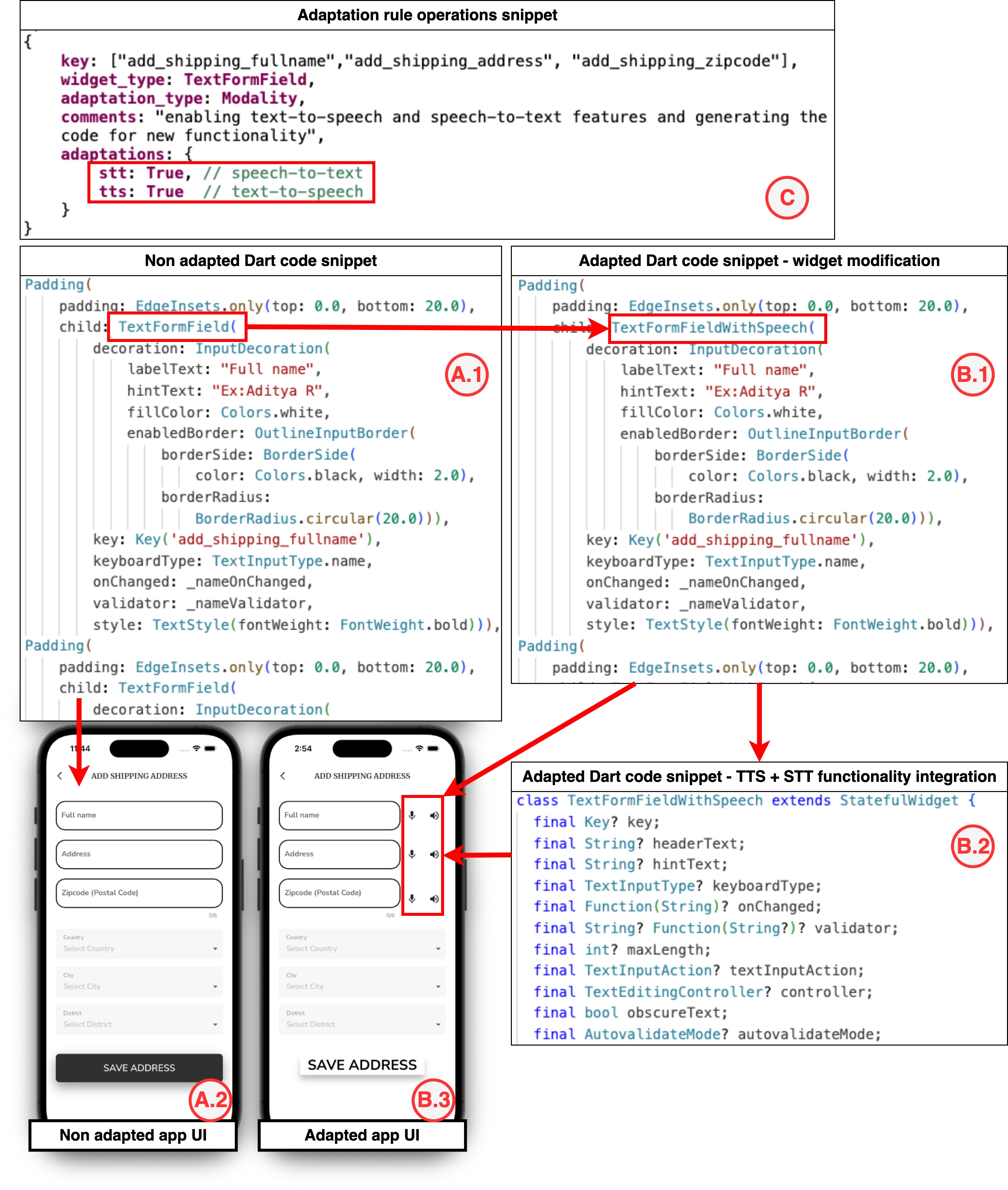}
\caption{An example of multi-modality adaptation with source code transformation. The previous presentation adaptation example is extended to integrate text-to-speech and speech-to-text functionality for the defined TextFormField widgets. In (C), when the speech-to-text and/or text-to-speech properties are set to true, the MDE pipeline identifies the relevant widget (A.1) and adapts the source code to include the necessary functionality, as shown in (B.1) and (B.2). This transformation is reflected in the app’s front end, where the adaptation evolves from instance (A.2) to (B.3).}\label{multimodality adaptation example}
\end{figure}

Adapting the Dart source code to add text-to-speech and speech-to-text functionality is more complex compared to presentation adaptations. It involves generating the required imports (such as Flutter's Text-to-Speech and Speech-to-Text plugins) and the corresponding Dart code to produce an executable UI. However, this task is feasible through the proposed approach, as demonstrated by the source code example in Figure \ref{multimodality adaptation example}.

\subsubsection{Navigation adaptation}

\textcolor{black}{Compared to the previously demonstrated adaptation types, navigation adaptations are arguably the most complex. When senior users require modifications to an application's navigation or workflow, the resulting changes to the source code can be highly variable. This variability makes it challenging to define a standard implementation method. One accessibility concern identified through our exploratory and evaluative focus group studies is that seniors often prefer less overwhelming user interfaces, particularly when experiencing age-related cognitive limitations \cite{wickramathilaka2025guidelines}. Moreover, certain adaptations, such as increasing UI element sizes or incorporating additional modalities like audio, can inadvertently introduce new accessibility barriers by increasing vertical scroll space or overall interface complexity.}

\textcolor{black}{To address this issue, a promising strategy is to segment UI content and present it incrementally, thereby simplifying the user's interaction workflow \cite{ahmad2020, wickramathilaka2025guidelines}. This type of adaptation alters the app’s navigational structure, and we selected it as a use case to demonstrate how our approach manages the code generation challenges typically associated with such workflow modifications.}

\begin{figure}
\centering
\includegraphics[width=1\textwidth]{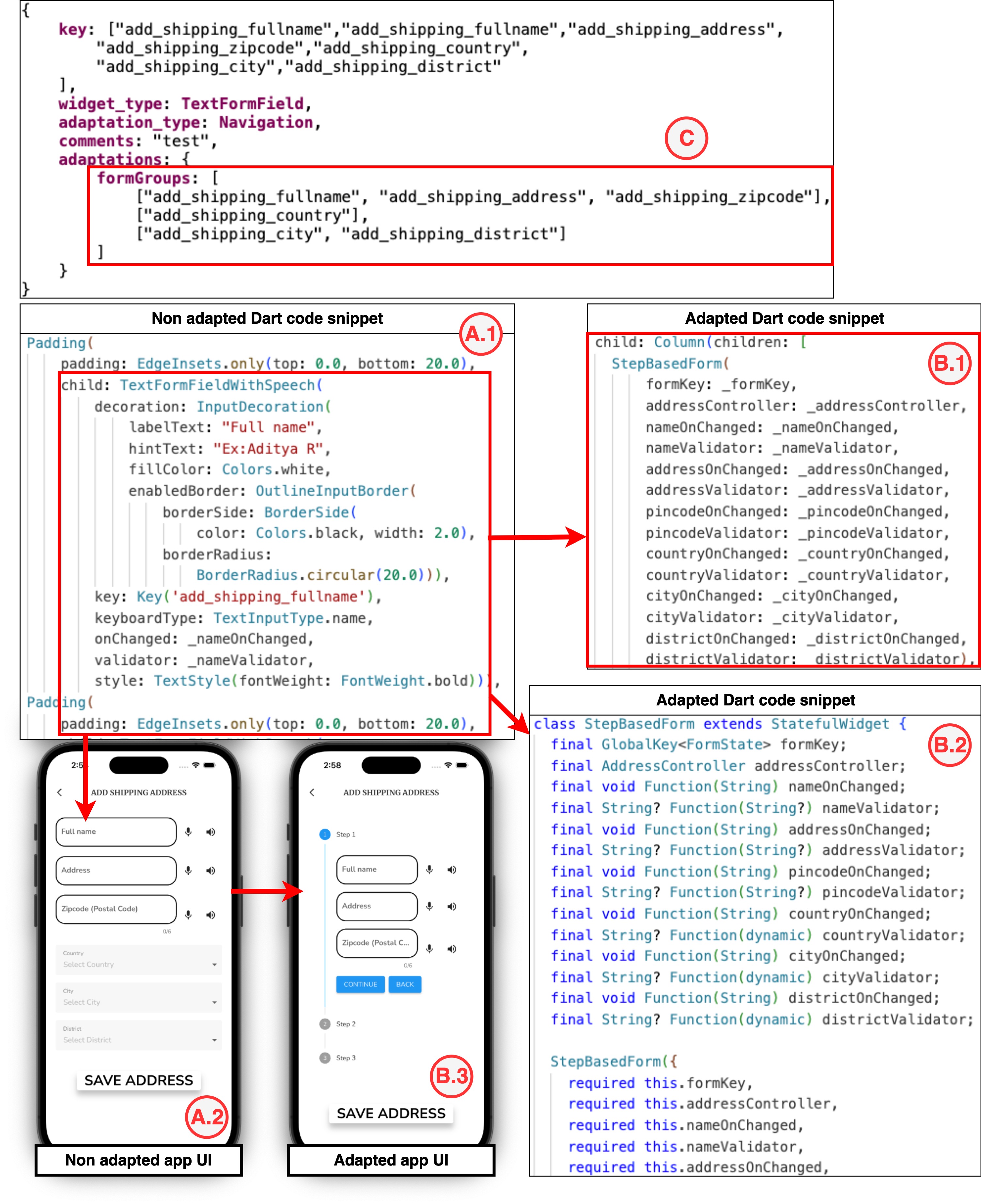}
\caption{An example of navigation adaptation with source code transformation. The previously adapted UI source code (A.1) is further transformed into a segmented form through automatically modified source code (B.1). (B.2) shows the additional classes and functionalities applied to the original code to enable this operation. To trigger this transformation, the developer defines the grouping and ordering of the relevant TextFormField widgets through the adaptation rules model (C). The final transformation to the UI instance is shown, evolving from (A.2) to (B.3).}\label{navigation adaptation example}
\end{figure}

\textcolor{black}{Accordingly, we used DSL inputs to dynamically introduce a wizard-style interface into an originally static, single-page form. This allows developers to restructure a single-page form into a dynamic stepper form, with the segmentation of form fields defined through our DSL. Figure~\ref{navigation adaptation example} illustrates how \toolname~achieves this adaptation. Unique widget identifiers from the original form (A.2) are referenced in the adaptation rule to specify three form groups. In the adapted UI (B.3), the MDE pipeline generates source code that separates the original form into these three groups, as specified in the rule's configuration.}

% However, we have demonstrated an example of form-based navigation adaptation using our proposed tool suite and approach, which minimises the need for manual coding by the developer, as illustrated in Figure \ref{navigation adaptation example}. In this example, a form with multiple input fields is transformed into a step-by-step form, allowing a senior user to sequentially navigate through a complex form in a mobile application. Using our Adapt DSL specification, we can dynamically group and order the input fields in the form, requiring the developer to add only a single line of code to the original app source code (a unique key property as a widget identifier).

\subsubsection{Comment generation}

In addition to the adaptation types discussed above, we provide support to developers in the specification and maintenance of comments for each adaptation rule and its associated granular adaptation operations (shown in Figure \ref{rules model example}). This allows developers to document the purpose of each rule and the specific adaptations it addresses, particularly concerning the senior user's age-related needs. Notably, this documentation can include references to the relevant WCAG guidelines that inform the adaptation rule. When the MDE pipeline modifies the source code, it embeds these comments within the modified Dart files. This practice provides developers with a clear documentation trail, helping them track any adaptations made to the original source code.

\subsection{Implementation}

To develop our prototype \toolname~tool suite, we utilized the Eclipse Modelling Framework (EMF) and its Domain-Specific Language (DSL) workbench tools. For the Context DSL, we implemented the DSL using the Sirius workbench \cite{sirius}. Additionally, we employed Acceleo \cite{acceleo} as a model-to-text transformer to convert the Sirius run-time model instance into a cleaner XML format. The development of the Adapt DSL was conducted using Eclipse's Xtext framework \cite{xtend}, with Xtend serving as the model-to-text transformer to generate a JSON structure from the Adapt DSL's textual run-time model instances.
The current prototype tool suite is also tightly integrated with the Flutter framework. Specifically, Flutter's widget documentation was used to define the widget attribute adaptation classes for the Adapt DSL. In its current implementation, the MDE pipeline exclusively generates and adapts Dart code.

\FloatBarrier

\section{Evaluation} \label{evaluation}

We evaluated \toolname\ using two separate but parallel qualitative user studies. Our approach's primary target audience was software developers. Therefore, we aimed to present \toolname\ in action (as a video demonstration) and gather their feedback regarding its usefulness, practicality, ease of use, potential for real-life adoption, and suggestions for improvement. On the other hand, the ultimate beneficiaries of our study are seniors. Thus, our second study -- a focus group study focused on demonstrating the adaptive UI prototypes and obtaining feedback from seniors to assess whether their age-related accessibility needs have been effectively addressed. Both studies were approved by the Monash Human Research Ethics Committee (MUHREC Project ID: 42470).

This paper primarily focuses on the feedback from software developers, as they are the intended users of \toolname. However, developers will not use the tool in isolation, without considering its impact on end-users. Therefore, we also provide a summary of the feedback received from seniors regarding the UI prototypes generated using \toolname.

\subsection{Evaluation with developers}

\subsubsection{Study protocol}

In this study, we opted to do 1-hour interviews with software developers. Due to the complexities and familiarity necessary to install and set the prototype in a participant's own device, we adopted an approach where we simulated an adaptation workflow with \toolname~running on the first author's personal computer and produced three video demonstrations\footnote{\url{https://drive.google.com/drive/folders/11b_P07ybFD7SoDS1r7rrdWd7wks1la6n?usp=sharing}}.

At the beginning of the interview session, we presented a scenario where a significant portion of a Flutter app's user base was aged above 60. Then we explained how these senior users find accessibility barriers in the app due to their diverse and personalised age-related accessibility needs and our proposal to address this.

In the first video demonstration, we used a simplified representation of Figure \ref{persona example}, which explained Judy's accessibility needs. We then used the Context DSL editor to create a context-of-use model representation of Judy. After the demonstration, we asked questions from the participants with regards to how practical and useful they found the modelling tool, its ease of use, suggestions for improvement, and the likelihood of the participant adopting the modelling tool in a real-life software development project.

In the second demonstration, we continued the same example by defining an Adaptation Rules Model for Judy's accessibility scenarios via Adapt DSL's editor. We then input both the context-of-use model and adaptation rules model into the MDE pipeline and generated an adapted app instance based on them. These adaptations were solely focused on presentation aspects (e.g., making text larger and bolder, changing the theme to black and white, enhancing the input box and button borders etc...). Again a similar set of questions to the previous demonstration were asked from the participant.

The final demonstration was focused on applying complex adaptations such as the integration of text-to-speech and speech-to-text features into the UI and transforming a traditional form into a dynamically grouped wizard pattern-based step-by-step form. We wanted to determine the practical implication of such complex source code adaptations where the source code differs drastically compared to much simpler presentation adaptations.

We then concluded the session where we invited the participant to express their opinions on their overall impressions and the \toolname's potential for adoption in a real-life software development project.

\subsubsection{Participant recruitment}

We recruited 18 participants to the study. Our primary recruitment criteria were for the software developers to have at least 6 months of professional experience in the Flutter framework. The recruitment was done through social media advertisements and authors' personal and professional contacts. Participation in this study was voluntary.

\subsubsection{Data collection}

Each interview session was carried out over Zoom, except for one that was done in person. \textcolor{black}{Both Zoom sessions and the in-person session followed the same interview protocol, and therefore, how the interview was conducted does not have any impact on results.} Session durations ran from 45 minutes to 85 minutes. The average duration per session was 59.9 minutes. We audio-recorded the sessions via Zoom and transcribed the data with Otter.ai.

\subsubsection{Data analysis}

After generating the interview transcripts, we carefully cleaned the dataset and familiarised ourselves with it. Subsequently, we subjected the data to a thematic analysis process. The initial coding revealed a diverse set of concepts, requiring iterative refinements to eliminate redundancies and streamline the codes into a hierarchical structure. Ultimately, we classified the codes under the following root themes: (1) Positive Reactions from Developers for \toolname, (2) Negative Reactions from Developers for \toolname, and (3) Suggestions for Improving \toolname.

While classifying the codes under these overarching themes was intuitive, identifying sub-themes within them proved challenging due to the variability in developer feedback. For instance, under the Suggestions for Improving \toolname~theme, developers offered a highly diverse range of suggestions. Nevertheless, we successfully organised the codes into a more manageable, theme-based hierarchy by the end of the analysis.

\subsubsection{Participant information}

\begin{table}
    \centering
    \renewcommand{\arraystretch}{1.2} % Adjust row spacing
    \resizebox{1\textwidth}{!}{
    \setlength{\tabcolsep}{5pt} % Adjust column spacing
    \small % Reduce font size if needed
    \begin{tabular}{|l|>{\raggedright\arraybackslash}p{1.5cm}|>{\raggedright\arraybackslash}p{2cm}|>{\raggedright\arraybackslash}p{2cm}|>{\raggedright\arraybackslash}p{2cm}|>{\raggedright\arraybackslash}p{2cm}|>{\raggedright\arraybackslash}p{3cm}|}
        \hline
        \textbf{ID} & \textbf{Age} & \textbf{Software industry exp.} & \textbf{Mobile dev. Exp.} & \textbf{Flutter exp.} & \textbf{Use accessibility standards?} & \textbf{Importance of addressing age-specific UI needs} \\ \hline
        D1  & 21 - 34  & 5 - 10 years  & 2 - 4 years  & 2 - 3 years  & Yes    & Important  \\ \hline
        D2  & 21 - 34  & 5 - 10 years  & \texttt{<} 1 year  & \texttt{<} 1 year  & Yes    & Essential  \\ \hline
        D3  & 21 - 34  & 2 - 4 years   & \texttt{<} 1 year  & \texttt{<} 1 year  & Yes    & Essential  \\ \hline
        D4  & 21 - 34  & 5 - 10 years  & 5 - 10 years  & 2 - 3 years  & Yes    & Essential  \\ \hline
        D5  & 21 - 34  & 2 - 4 years   & \texttt{<} 1 year  & \texttt{<} 1 year  & Yes    & Essential  \\ \hline
        D6  & 21 - 34  & 5 - 10 years  & 5 - 10 years  & \texttt{>} 5 years  & Yes    & Important  \\ \hline
        D7  & 21 - 34  & 5 - 10 years  & 2 - 4 years  & 2 - 3 years  & Maybe  & Important  \\ \hline
        D8  & 21 - 34  & 2 - 4 years   & 2 - 4 years  & 2 - 3 years  & Maybe  & Important  \\ \hline
        D9  & 21 - 34  & \texttt{<} 1 year  & \texttt{<} 1 year  & \texttt{<} 1 year  & Yes    & Essential  \\ \hline
        D10 & 21 - 34  & 5 - 10 years  & 5 - 10 years  & \texttt{>} 5 years  & Maybe  & Important  \\ \hline
        D11 & 21 - 34  & 2 - 4 years   & 2 - 4 years  & 3 - 4 years  & Maybe  & Essential  \\ \hline
        D12 & 35 - 59  & \texttt{>} 10 years    & 2 - 4 years  & 3 - 4 years  & Yes    & Important  \\ \hline
        D13 & 21 - 34  & \texttt{<} 1 year  & \texttt{<} 1 year  & \texttt{<} 1 year  & No     & Somewhat Important  \\ \hline
        D14 & 21 - 34  & 5 - 10 years  & 5 - 10 years  & 3 - 4 years  & No     & Important  \\ \hline
        D15 & 35 - 59  & \texttt{>} 10 years    & \texttt{>} 10 years   & \texttt{>} 5 years   & Yes    & Essential  \\ \hline
        D16 & 35 - 59  & 5 - 10 years  & 5 - 10 years  & 2 - 3 years  & Maybe  & Somewhat Important  \\ \hline
        D17 & 21 - 34  & 5 - 10 years  & 5 - 10 years  & 3 - 4 years  & Yes    & Somewhat Important  \\ \hline
        D18 & 21 - 34  & \texttt{>} 10 years  & 2 - 4 years  & 3 - 4 years  & No & Somewhat Important  \\ \hline
    \end{tabular}
    }
    \caption{Demographic data of software developers}
    \label{tab:developer_data}
\end{table}

\begin{table}
    \centering
    \renewcommand{\arraystretch}{1.3}
    \resizebox{\textwidth}{!}{
    \begin{tabular}{|p{2cm}|*{6}{>{\centering\arraybackslash}p{2cm}|}}
        \hline
        \textbf{Developer ID} & \textbf{Design universally accessible UI based on accessibility guidelines} & \textbf{Integrate native accessibility settings} & \textbf{Capture personalisation needs via personas} & \textbf{Provide configuration dashboard for UI personalisation} & \textbf{Use general practices unless special case handling is needed} & \textbf{Rarely personalise UI} \\ \hline
        \textbf{D1} & X &  & X & X &  &  \\ \hline
        \textbf{D2} & X & X & X &  &  &  \\ \hline
        \textbf{D3} & X & X & X &  &  & X \\ \hline
        \textbf{D4} & X & X &  & X &  &  \\ \hline
        \textbf{D5} &  & X & X & X &  &  \\ \hline
        \textbf{D6} & X & X &  & X &  &  \\ \hline
        \textbf{D7} & X & X &  &  &  &  \\ \hline
        \textbf{D8} &  &  &  &  &  & X \\ \hline
        \textbf{D9} &  &  & X & X &  &  \\ \hline
        \textbf{D10} &  &  &  &  &  & X \\ \hline
        \textbf{D11} & X &  &  &  &  &  \\ \hline
        \textbf{D12} &  & X &  &  &  &  \\ \hline
        \textbf{D13} & X &  & X &  &  &  \\ \hline
        \textbf{D14} & X & X &  & X &  &  \\ \hline
        \textbf{D15} & X & X & X & X &  &  \\ \hline
        \textbf{D16} &  & X &  &  &  & X \\ \hline
        \textbf{D17} & X &  &  &  & X &  \\ \hline
        \textbf{D18} & X & X &  &  &  &  \\ \hline
        \textbf{\%} & \textbf{66.7\% (12)} & \textbf{61.1\% (11)} & \textbf{38.9\% (7)} & \textbf{38.9\% (7)} & \textbf{5.6\% (1)} & \textbf{22.2\% (4)} \\ \hline
    \end{tabular}
    }
    \caption{Current strategies used by developer participants to address age-specific UI needs}
    \label{tab:personalisation_strategies}
\end{table}

The demographic data for our evaluation study with software developers (N=18) is provided in Table \ref{tab:developer_data}. In terms of age distribution, the majority of participants (n=15, 83.3\%) were between 21–34 years, while the remaining n=3 (16.7\%) fell within the 35–59 age category. We then assessed participants' experience levels across three key dimensions: (1) overall professional experience as developers, (2) experience in mobile development, and (3) experience in the Flutter framework.

Regarding overall professional experience in the software industry, most developers (n=9, 50\%) reported having 5–10 years of experience. Additionally, three participants (16.7\%) had more than 10 years of experience, while two (11.1\%) were novice developers with less than one year of experience. With respect to experience in mobile app development, the levels varied; however, only a minority (n=5, 27.8\%) reported having less than one year of experience. A similar trend was observed for experience in the Flutter framework, where again, five participants had less than one year of experience. Notably, we were able to recruit developers with 3–4 years (n=5, 27.8\%) and even more than five years of experience in the framework (n=3, 16.7\%) -- an encouraging finding, given that Flutter 1.0 was only released in late 2018 \cite{flutter}.

Next, we aimed to assess the current priorities and strategies our developer participants follow when developing accessible software, particularly for senior users. First, we inquired whether participants incorporate accessibility standards, such as WCAG, into their development practices. Encouragingly, 10 participants (55.6\%) reported that they do. To further explore developers' perspectives on addressing accessibility barriers in UIs for seniors, we asked them to rate the importance of considering senior age-related accessibility needs using a Likert scale. Notably, 7 participants (38.9\%) stated that it is essential, while another 7 participants (38.9\%) considered it important. This indicates that the vast majority of our sample acknowledged the significance of accessibility considerations in UI development.

The final question in our pre-interview questionnaire (Table \ref{tab:personalisation_strategies}) aimed to gain a concrete understanding of the exact strategies software developers employ when designing UIs and apps for senior users. The most prevalent strategy among participants (n=12, 66.7\%) was designing universally accessible UIs with accessibility standards such as WCAG in mind. Another commonly adopted approach (n=11, 61.1\%) was ensuring that UIs comply with native operating system-level accessibility settings and requirements. Encouragingly, a majority of participants (n=11, 61.1\%) reported having strategies to support personalisation for their users. Specifically, seven participants (38.9\%) stated that they interact with end users to create personas, while seven of them (38.9\%) indicated that they provide a UI configuration dashboard for end-user customisation. Conversely, a minority (n=4, 22.2\%) admitted that they rarely take steps to personalise their UIs for seniors, citing time and resource constraints as limiting factors.

\subsubsection{Results}

A key theme that emerged from our analysis was how \toolname\ could either enhance or hinder the developer experience in a software development setting. Nearly all participants (n=17) provided positive feedback on the strengths of \toolname, highlighting how it could make the development process more intuitive and less burdensome when designing personalised software UI for seniors and other user groups. However, some participants (n=10) also expressed concerns, offering critical feedback on potential drawbacks. We identified that these limitations stem from the current implementation of the prototype, which can be improved in future iterations based on the improvement suggestions provided by all participants (n=18). In this section, we examine and summarise the key findings of the study, encompassing positives, negatives, and suggestions for improvement as discussed by the majority of participants.

\subsubsection*{Developer feedback on \toolname's strengths}

\textbf{Ease of use in DSLs:}

\textcolor{black}{The design and usability of the DSLs received highly positive feedback from developers, with 17 out of 18 interviewees expressing approval.} Individually, 15 developers found Context DSL user-friendly, while Adapt DSL was similarly well-received (n=15). For Context DSL, developers particularly appreciated its tree-like hierarchy, as it aligns with their existing mental models. As [D2] noted, \textit{"I think it's very easy to understand because it's very well structured, and it's like a tree, so it's very easy for developers to understand, because, you know, the structure just makes sense itself."} \textcolor{black}{Similarly, Adapt DSL was praised for its user-friendliness, with developers finding its JSON-like structure intuitive and easy to understand. As [D7] mentioned, \textit{"It's very user friendly due to its JSON structure, and we all know JSON, and it's like, not [a] very steep learning curve"}}

\noindent\textbf{Enhanced developer convenience:}

Another recurring theme (n=11) in the data regarding developers' experience with \toolname\ was its potential to provide convenience in addressing age-related needs for end-users. Developers were particularly positive about its automated code generation capabilities through DSL models, which they felt could save time and reduce effort compared to traditional development methods. For example, [D5] noted, \textit{"I think it saves a lot of time for a developer, because, in order to deal with all the adaptations happening in the user interface. Especially, when you have a lot of users with different needs."}

\noindent\textbf{Usefulness and practicality of the approach:}

17 out of 18 participants agreed that our proposed \toolname\ tool and its subcomponents are useful and beneficial in achieving our goal: finding a better approach to addressing age-related accessibility needs in UIs for seniors. Even [D14], who was the most critical participant in our study, acknowledged that at least one of our subcomponents, Context DSL, could be situationally useful. This is further evidenced by what was said by [D10] regarding the overall usefulness of \toolname~, \textit{"I'm really interested in using this kind of tool if I'm developing this kind of app, and I think it's really useful and it can save a lot of developers' time, and it will make the app more usable to the senior users as well."}

In addition, a majority of the participants (n=16) stated that the approach that we have taken in this project is a practical one. Of course the developers had a wide range of improvement suggestions to ensure that our next tool iterations mitigate the practical considerations raised regarding our current prototype implementation. For example, the following was noted by [D15]: \textit{"Outside my previous suggestion of generated files, I'm pretty impressed. And I think it could be a maintainable solution. I think it's accessible for developers."}

\vspace{1em}
\noindent\textbf{Potential for real-life adoption:}

We specifically asked our study participants whether they would be willing to use our proposed tool and its subcomponents in a real-life UI development project, particularly when designing for a user base that includes a significant proportion of senior users. The feedback we received was overwhelmingly positive, with 16 out of 18 participants agreeing that they would be inclined to adopt our approach. For example, [D1] noted, \textit{"I would definitely use it. And like I said, this might be a must-use model or approach in the future."}. Another participant -- [D7] told us, \textit{"I find it very useful, and I'm kind of impressed. And at the same time, I was wondering why we don't have such a tool at the moment. And I'm really happy that people like you are doing such implementations, and I think it will help 1000s and 1000s of developers, plus millions of users, while using and while implementing, both scenarios to have this kind of tools."}

\subsubsection*{Developer feedback on \toolname's weaknesses and potential mitigation strategies}

\noindent\textbf{Incurring extra effort:}

A non-majority (n=7) of participants raised a concern about how \toolname~ could require additional effort from developers. For example, [D10] highlighted the extra effort required to learn and integrate the tool into existing development workflows, stating, \textit{"It takes some extra time when we are implementing this, because this is a totally separate thing. ...... Currently, since we are not making it [app] dynamic, it's mostly static. So if you are making it dynamic, it will take some extra time."} A different but solitary concern was from {D14}, where they pointed out the fact that considering age-related needs in an app might fall outside a developer's typical responsibilities: \textit{"It's a UX guy job. It's not a developer's job to do all of this, right?"}

\underline{\textit{Mitigation strategies:}} A significant portion of our participants (n=8) proposed that we need mechanisms in place to provide developers with guidance when it comes to modelling tasks. Concrete examples, personas, and model documentation metadata could help developers in reducing the learning effort. This is evidenced by [D8] who stated, \textit{“I think one thing that could be better is if we have recommended way of doing something right, so without just giving the operations you can just set up, okay, these are the conditions and then allow users to use that as a template to begin with, so they know what sort of like combinations we how to consider to start with”}.

\noindent\textbf{Infeasibility in collecting users' context data:}

A minority of the participants (n=5) commented on how impractical it would be to capture the user, platform, and context parameters of the users in order to facilitate our proposed approach in the first place. To illustrate this challenge, [D4], a UX engineer, highlighted concerns about the onboarding process and user retention, stating, \textit{"How long will it take to configure these things when they [users] install the application? Because that's also a concern. Sometimes, when you are creating [an] application, [it] is much more concerning about how speedy the onboarding process is. If it is too complicated for any use, it doesn't matter that like age [related needs] or like if they need personalised usage, but if we make it a little longer or [add] extra steps, it can be a concern of the user. Like, it can drop user from the installing the app, or drop [them] out [from] doing some actions in the application."}

\underline{\textit{Mitigation strategies:}} Four interviewees provided suggestions on this matter and we identified two potential strategies to resolve it. (1) Seniors volunteering the accessibility information during the onboarding process: This is the more traditional approach and was proposed by both [D12] and [D15]. Despite [D4]'s doubts about putting a burden such as this on the seniors leading to a drop in user base, [D15] is of the opinion that it is a viable approach as evidenced by his statement: \textit{"My mom would definitely use it, for sure. She already uses some of those, like text scaling and stuff like that on her phone"}; and (2) A universal accessibility profile: This method was suggested by both [D8] and [D12] and the latter had a great example to showcase its approach: \textit{"Hey, look, you can create yourself a profile of your impairments over here. And then apps can subscribe to that service so that when that person, when Judy, logs into this new app, she needs to be able to do the New Zealand travel declaration, the New Zealand traveller declaration will go over to that service, grab her profile and then apply it all to the UI, yeah? So she doesn't have to do that for every single app"}

\noindent\textbf{Code scalability, maintainability and versioning:}

Developing even a small app would take tens of thousands of code lines. For example, [D14] shared with us that an app he's developing for a small startup reached over 80,000 lines of code. Even in such cases, where the user base is relatively small, implementing personalized UI adaptations using our proposed approach to adaptive code generation could significantly increase the codebase size. With the need to store multiple adaptive permutations, the total number of lines of code could grow exponentially, leading to both code readability and maintainability issues, as noted by a group of participants (n=6). For example, [D14] noted, \textit{"the second a human needs to look at it and fix it and change it [codebase], we have to have some kind of maintainability or structure or readability"}. Consequently, this issue also leads to issues with version control. As explained by [D10], \textit{"So definitely, we are using something like GitHub or GitLab. So you have to put the code there if you are doing a change again, like UI, change functionality, .... you have to add all these new changes to each version."}

\underline{\textit{Mitigation strategies:}} To prevent the codebase from becoming unwieldy or unmaintainable due to the source code adaptations performed by \toolname, participants suggested several strategies. One notable approach was: Slotting in widgets based on API calls. In this method, the application is designed so that its source code includes allocated spaces for any widgets that may need adaptation. When an adaptation is triggered, the necessary operations and context parameters are sent to an adaptation service REST API, which then serves the required adapted widgets. These widgets are dynamically slotted into the source code, creating an adapted app instance without requiring complex code generation. For example, [D10] described this approach as follows: \textit{"In the Flutter you provide the space. So that provided space, you'll be listening to the back end again. And with the adaptation coming, ..... you're replacing the widget with the new widget coming according to the adaptation from the back [end]"}. A key advantage of this approach is that developers would no longer need to maintain multiple code repositories for each app permutation, reducing code duplication and simplifying maintenance.

\vspace{3em}
\noindent\textbf{Deployment concerns:}

Another practical issue raised by participants (n=5) was the deployment of adaptations to end-users in a real-world setting. For example, [D17] explained to us that they optimise the app bundle size so that the app is able to work efficiently even on an older mobile device that has limited resources such as storage. This leads to a challenge where it would be disadvantageous for us to store an adaptation service, DSL models, and conditionally generated code locally. Thus, it would require us to maintain a centralised server, allowing over-the-air (OTA) updates to deliver necessary adaptations to users' apps. However, OTA-based adaptations face significant restrictions under current App Store and Play Store policies. As [D15] pointed out, \textit{"The question of it being in an App Store is kind of a problem, right? Because you have to go through a lot of like getting approvals for app changes."} Even minor updates or bug fixes require going through a lengthy and unpredictable approval process, making frequent or dynamic adaptations impractical within the existing mobile app ecosystem.

\underline{\textit{Mitigation strategies:}} To mitigate the limitations imposed by app store policies on providing updates for UI adaptations, we can leverage tools such as Shorebird or similar services to push over-the-air (OTA) updates to the app almost instantly. This approach was explained by [D12] during our interview: \textit{"The only way you could potentially do it and have the code somewhere else is, I don't know if you've seen 'Shorebird', it does over-the-air updates, so you release your app to the store, and then you can do patches, and you don't have to go through the store review process to get the patch. So you could potentially have a modified version of Shorebird that allows you to pull particular code for those data adaptations."} 

\subsection{Evaluation with seniors}

\subsubsection{Study protocol}

We conducted a focus group study to gather feedback from participants. A presentation-based approach was used to present the questions and demonstrate the application using short video clips\footnote{\url{https://drive.google.com/drive/folders/1_Eooe_XxzAHi8cfUHfWHP8WB4SyP55fs?usp=sharing}}. At the beginning of the session, a Flutter application was shown to participants through a short video clip. The application itself is open-source\footnote{\url{https://github.com/adeeteya/FlutterFurnitureApp}} and is a generic retail furniture app. The demonstration was originally performed on an actual mobile device, where we screen-recorded a user logging in, browsing furniture products, and navigating to a form page to add a shipping address.

The focus group session was structured around three types of adaptations: presentation, multi-modality, and navigation, with each section beginning with a video demonstration. First, participants viewed the app in its original, non-adapted form and provided feedback on usability and accessibility challenges. The second demonstration focused on presentation adaptations, applying \toolname\ to enhance text size, input box borders, and contrast. Participants compared this adapted version with the original. The third demonstration introduced multi-modality adaptations, adding text-to-speech and speech-to-text features for the shipping address form. The fourth showcased a navigation adaptation, transforming the multi-field form into a step-by-step wizard. The session concluded with a discussion on app run-time adaptations to inform future work.

\subsubsection{Participant recruitment}

We recruited 22 senior participants from a University of the Third Age (U3A) chapter in Australia. Each participant was gifted a 30 AUD voucher in a local supermarket as a token of appreciation.

\subsubsection{Data collection}

We held three focus group sessions in person and each ran approximately for an hour. Each session was audio recorded and then transcribed using Otter.ai. During data collection, we inquired about any accessibility barriers or enablers that participants encountered while using the prototype UIs.

\subsubsection{Analysis}

After carefully cleaning the qualitative data transcripts collected from the focus groups, we conducted a thematic analysis. During the initial coding process, the data revealed a diverse set of codes. Through an iterative refinement process, we organised these codes into three overarching root themes: (1) Enablers to Accessibility, (2) Barriers to Accessibility, and (3) Factors Influencing Personalised User Experiences.

The first two themes effectively summarised the insights shared by our senior participants regarding their current experiences with app UIs. These themes captured whether participants found the proposed personalisation methods (demonstrated through UI prototypes) useful and how they envisioned further enhancing their app experiences.

In contrast, the third theme focused on how seniors perceive personalising their app UIs through their own user agency. Within this theme, we further classified the codes into enablers and disablers for personalisation, as well as explored seniors’ perceptions of the methods through which they exercise their personalisation agency. However, this paper reports only on the first two themes, as they focus on feedback about the current version of our prototype, while the third theme will inform future work.

\subsubsection{Participant information}

The demographic information of the senior participants in our evaluation study is provided in Table \ref{tab:senior_demographic_data}. The mean age of the participants was 72.1 years and the median age was 72 years. The majority were women (n=17, 77.3\%), and most identified culturally as Australians (n=14, 63.6\%). Participants had a diverse range of prior occupations, reflecting varied skills, experiences, and backgrounds.

Nearly all participants used spectacles as a visual aid (n=21, 95.5\%), indicating common age-related vision impairments. Additionally, all participants owned a mobile device, with the majority also owning a tablet. The number of hours per week spent using mobile apps varied, but half of the participants (n=11. 50\%) reported using them for more than 10 hours per week. Most participants (n=13, 59.1\%) indicated an average confidence level in using mobile apps, while only two participants (9.1\%) reported below-average confidence.

\begin{sidewaystable}
    \centering
    \footnotesize
    \begin{tabular}{|l|ll>{\raggedright\arraybackslash}p{2cm}>{\raggedright\arraybackslash}p{2cm}|>{\raggedright\arraybackslash}p{2cm}|l|>{\raggedright\arraybackslash}p{2cm}|>{\raggedright\arraybackslash}p{2cm}|}
        \hline
        \textbf{\centering ID} & \textbf{Age} & \textbf{Gender} & \textbf{Culture} & \textbf{Occupation (prior)} & \textbf{Usage of aids} & \textbf{mobile device usage} & \textbf{App usage (hours-per-week)} & \textbf{Mobile app usage confidence level} \\ \hline
        P1  & 74 & F & Australian & Public servant & Eye glasses & Android phone, Tablet & 4-9 & Average \\ \hline
        P2  & 85 & M & Anglo & Engineer & Eye glasses, Hearing aids & - & \texttt{<}1 & Very low  \\ \hline
        P3  & 75 & M & Australian & - & Eye glasses & iPhone, Tablet & 10-20 & high  \\ \hline
        P4  & 72 & F & Australian & Nurse & Eye glasses & Android phone & 4-9 & Average  \\ \hline
        P5  & 76 & F & Indian & Educator, Academic & Eye glasses & iPhone, Tablet, Kindle & 2-3 & Average  \\ \hline
        P6  & 68 & F & English & Academic librarian & Eye glasses & iPhone, Tablet & 4-9 & Very high  \\ \hline
        P7  & 69 & F & Australian & Dental assistant & Eye glasses & Android phone & 20\texttt{<} & Average  \\ \hline
        P8  & 72 & F & Australian & Nurse & Eye glasses & iPhone, Tablet & 10-20 & Average \\ \hline
        P9  & 83 & M & Australian & Business owner & Eye glasses, Hearing aids & Android phone & \texttt{<}1 & Average  \\ \hline
        P10 & 66 & F & Australian & Marketing & Eye glasses & iPhone & 20\texttt{<} & High  \\ \hline
        P11 & 63 & F & Dutch & Community support & Eye glasses & Android phone & 10-20 & Average  \\ \hline
        P12 & 62 & F & Chinese & - & n/a & Android phone, Tablet & 2-3 & High  \\ \hline
        P13 & 75 & F & Australian & Casual librarian & Eye glasses & iPhone, Tablet & 20\texttt{<} & High  \\ \hline
        P14 & 80 & M & German & Industrial sales & Eye glasses, Hearing aids & iPhone, Tablet & \texttt{<}1 & Average \\ \hline
        P15 & 68 & F & Australian & Trainer, Assessor & Eye glasses & iPhone & 10-20 & Average to low \\ \hline
        P16 & 61 & F & Australian & Information management officer & Eye glasses & Android phone, Tablet & 10-20 & Average  \\ \hline
        P17 & -  & F & Indian & Admin officer & Eye glasses & Android phone & 4-9 & Average \\ \hline
        P18 & 70 & F & Australian & Banking & Eye glasses & Android phone, Tablet & 10-20 & Average  \\ \hline
        P19 & 68 & M & Australian, Irish & Public servant & Eye glasses & Android phone & 2-3 & High \\ \hline
        P20 & 79 & F & Australian & Funeral director & Eye glasses & Android phone, Tablet & 20\texttt{<} & Average  \\ \hline
        P21 & 76 & F & Australian & Service assistant & Eye glasses & iPhone, Tablet & 20\texttt{<} & Average \\ \hline
        P22 & 73 & M & Scottish & Senior manager IT & Eye glasses & Android phone, Tablet & 2-3 & High  \\ \hline 
    \end{tabular}
    \caption{Senior participant demographic information. N = 22 seniors participated in the study.}
    \label{tab:senior_demographic_data}
\end{sidewaystable}

\subsubsection{Results}

Across all three focus groups, we received overwhelmingly positive feedback. Participants strongly resonated with the adaptive enhancements we proposed to make the demonstrated app more tailored to the senior population, particularly when compared to the generic, non-adapted app instances used as a baseline. In the following paragraphs, we summarise the findings regarding these enhanced features included in the adapted UI prototypes, categorised by their adaptation types: Presentation, Multi-modality, and Navigation. A more detailed description of our findings can be found in Wickramathilaka et al. \cite{wickramathilaka2025guidelines}.

\textbf{Presentation adaptations:} 
When we presented app UI instances with enhanced visual elements to our senior participants, we received positive feedback. For example, they expressed satisfaction with improvements to text readability, particularly the use of bolder and larger fonts. Another widely appreciated enhancement across all three focus groups (N=3) was the improved background-foreground contrast ratios. Beyond general text enhancements, we applied UI adaptations such as changing the app’s colour theme to black-and-white and adding enhanced borders around buttons, input fields, and dropdown menus. However, one concern that emerged was the trade-off between text visibility and the amount of information displayed on a mobile screen. Participants noted that this issue was not unique to our adaptations but was something they had encountered previously when adjusting text size either through their mobile device’s global font settings or using zoom-in gestures.

\textbf{Multi-modality adaptations:}
Our demonstrations on adapting app UIs with embedded alternative input and output features (e.g., Figure \ref{adapted app instances example}) were received very positively. Across all three focus groups, participants welcomed the option for seniors to use an alternative input method, specifically speech-to-text, instead of the traditional on-screen keyboard. To a lesser extent (in two out of three focus groups), participants also found text-to-speech features to be a helpful addition. Despite this positive feedback, seniors expressed scepticism about the viability of speech-to-text technology due to its limitations. A common theme across all focus groups was their prior experiences with inaccurate voice input, often attributed to the diversity in accents, dialects, intonations, and pronunciations. Additionally, the situational appropriateness of alternative input/output modalities was a key concern raised by the majority (two out of three focus groups). Participants noted that they would feel uncomfortable using audio-based features in public settings due to both privacy concerns and social etiquette.

\textbf{Navigation adaptations:}
The adaptations we demonstrated, which transformed a traditionally non-sequential, single-page form into a dynamic, wizard-pattern form, were highly appreciated by participants. A recurring theme across all three focus groups was that this approach made the presented information feel less overwhelming for seniors. Participants expressed that succinctly displaying information on-screen, combined with a logical and intuitive UI flow similar to a wizard-pattern form, was a clear improvement over the traditional static form layout typically used for senior end users.

\section{Limitations and Future Work} \label{limitations}

\subsection{Limitations}

\subsubsection{Limitations with \toolname}

Apart from the weaknesses noted by our developer study participants, another limitation of our current prototype is that the Adapt DSL is tightly coupled with Flutter's widget structure. To make the proposed approach more platform-agnostic, the Adapt DSL would need to be more abstract in specifying adaptation operations. Additionally, the code generation pipeline would have to be extended and rigorously tested to ensure that the UI adaptations align with the specific programming language or framework in use, ensuring flexibility across different platforms. 

Furthermore, more complex adaptation operations, such as senior requiring changes to the UI workflow of an app, may necessitate further extensions beyond what we have proposed. Seniors may have age-related needs that require adaptations to app functionalities such as dynamic page routing, customised workflow patterns, contextual help and guidance, multi-modal navigation support, and backward navigation or undo functionality. Implementing such adaptations would require considerable modifications to the app’s source code, making it difficult to automate these operations through DSL inputs in a platform- or framework-agnostic manner.

The issues mentioned above are further exacerbated by the inherent rigidity of Model-Driven Engineering (MDE) approaches compared to traditional software development tools. MDE pipelines are typically interconnected through multiple model-to-model and model-to-text transformations. As a result, additional care must be taken to ensure that these transformation processes are both flexible and robust, in order to avoid excessive "round-trip maintenance." Without careful design, this could hinder the ability to efficiently modify models and regenerate code. Ensuring flexibility in the transformation steps is critical to guarantee that the final output generates executable applications or functional code stubs without frequent manual intervention.

\subsubsection{Limitations with user studies}

Recruiting participants with at least six months of Flutter experience was challenging, likely due to Flutter's relative novelty, with its 1.0 release in late 2018 \cite{flutter}. Despite successfully recruiting 16 Flutter developers with varying experience levels, two participants ([D2] and [D3]) had familiarity with Flutter but not in a professional capacity. However, they were experts in developing software for seniors and well-versed in state-of-the-art accessibility guidelines such as WCAG \cite{wcag2.2} and ATAG \cite{atag}.

\textcolor{black}{Furthermore, demonstrations were conducted via video recordings to reduce the setup burden on our volunteer developer participants. Although the current prototype of \toolname~is functional, it has not yet been packaged as a streamlined Eclipse plugin. Instead, it requires manual setup of the Eclipse IDE, installation of the necessary Eclipse Modelling Framework (EMF)~\cite{emf} dependencies, and additional configurations for supporting tools such as Sirius~\cite{sirius}, Xtext~\cite{xtend}, Acceleo~\cite{sirius}, and Xtend~\cite{xtend}. Given this complexity, it would have been impractical to expect participants, especially those unfamiliar with Eclipse, to spend potentially over an hour configuring the environment before even using the tool. This expectation would have posed a significant barrier, considering their already demanding schedules as professional software developers. While the video demonstrations allowed us to communicate the key features and workflows of \toolname, they may not have captured the full depth of participant feedback. In particular, hands-on usage could have revealed nuanced insights regarding the tool's usability and areas for qualitative improvement in the overall developer experience.}

\subsection{Future work}

Our evaluation study with developers revealed many suggestions for improvements and feature additions. Our next iteration of \toolname~will especially focus on enhancements to code generation-related tasks, where we want to make our generated code more readable, maintainable, and reusable, mitigating the issues raised by our developer participants about the current prototype. In addition, many improvements were also proposed regarding our two novel DSLs, such as metamodel enhancements and model editor enhancements.

\textcolor{black}{With these improvements, we can package \toolname~as an Eclipse plugin with a streamlined installation process, accompanied by a supporting wiki to assist with configuration and setup. This enhancement would enable us to invite participants to engage in hands-on experimentation with the tool, allowing them to perform a range of tasks directly. Such an approach would facilitate the collection of richer, more nuanced feedback during the evaluation phase of our user study.}

Currently, we support only design-time UI adaptations. The next step is to extend these capabilities to enable user-requested run-time adaptations. Rather than pursuing automated self-adaptation, as proposed by Yigitbas et al. \cite{yigitbas2020}, we focus on empowering seniors to personalise their app UIs autonomously, ensuring the app does not dictate their needs. Our focus group evaluation study has already explored the barriers and enablers of run-time personalisation among seniors \cite{wickramathilaka2025guidelines}. To achieve this, we consider two approaches: a traditional accessibility settings-based method and a novel Large Language Model (LLM)-based approach, where LLMs transform multimodal natural language inputs into abstract DSL models, allowing us to reuse the MDE process in \toolname~to trigger app adaptations.

16 Meriton Pl, Clayton South

\section{Related Work} \label{related work}

% put this at end of paper - key things to compare to your approach - MDE solutions, adaptive UI solutions etc

Our study is primarily inspired by Yigitbas et al.’s \cite{yigitbas2020} work on an MDE-based approach for self-adaptive UIs, which also proposes two DSLs conceptually similar to ours. Their ContextML DSL models a user's context-of-use parameters, while AdaptML functions as a rules engine for conditionally applying UI changes. A key methodological difference is how UI changes are applied: while we modify the source code directly, Yigitbas et al. \cite{yigitbas2020} adopt an abstract UI model defined with the Interactive Flow Modelling Language (IFML) \cite{brambilla2014}. One limitation of their approach is the depth of modelling in their DSLs; for instance, ContextML lacks comprehensive support for age-related needs, such as UI preferences, and hearing and mobility impairments. Nonetheless, their study remains one of the most mature implementations in the MDE + Adaptive UI subdomain so far. We also appreciate that it is evaluated with real end-users, though it does not consider developers' feedback -- a crucial factor in understanding the enablers and barriers to real-world adoption by the software development community.

Another important study is by Bendaly Hlaoui et al. \cite{bendaly2018}, who propose an MDE approach for design-time UI adaptations tailored to disabled users. They achieve this transformation through two models: (1) an accessibility ontological instance representing an end user's context-of-use parameters and (2) a UI model reverse-engineered from an existing non-adapted UI. These models serve as inputs to an adaptation process that applies adaptation rules to transform the non-adapted model into an adapted one, which is then converted into the final UI. Their context-of-use modelling depth is well-developed and has informed the level of comprehensiveness needed for our study. However, it is unclear whether they produced an MDE prototype, as the authors mention the final transformation from an adapted abstract UI model to an executable UI as future work. Without this detail and a subsequent proof-of-concept evaluation, assessing the effectiveness of their reverse engineering approach remains challenging, particularly in terms of its compatibility with modern development tools and frameworks such as Flutter or React Native.

Minon et al. \cite{minon2015} propose an approach similar to ours, incorporating both context-of-use modelling and adaptation rules. Their tool, the Adaptation Integration System (AIS), automatically tailors UIs to meet the accessibility requirements of user groups with visual, hearing, and cognitive impairments at both runtime and design time. AIS includes a compilation of UI adaptation rules designed for these user groups. At design time, the MDE UI tool designer manually inputs a UI model at any abstraction level of the CAMELEON framework \cite{calvary2002}, along with parameters indicating the user’s disability. The system then generates an adapted UI model accordingly. However, the paper lacks detailed information about its metamodels, making it difficult to assess the depth of AIS’s modelling capabilities. The provided evidence, such as prototype applications and model examples, suggests that their proof-of-concept is less detailed than both ours and that of Yigitbas et al. \cite{yigitbas2020}. Additionally, unlike the studies mentioned above, Minon et al. \cite{minon2015} do not include any user or developer evaluation, limiting insights into its practical effectiveness.

A fundamentally different paradigm is proposed by Akiki et al. \cite{akiki2016} for MDE-based adaptive UI generation. They adopt a Role-Based UI adaptation mechanism that provides end-users with a minimal feature set and an optimal layout based on their context-of-use scenarios. In this approach, UI elements and adaptation rules are treated as accessibility resources and assigned to user accounts as ‘roles.’ At design time, a user’s account is allocated roles linked to context-of-use factors such as disabilities and culture. However, their modelling approach lacks sufficient depth to address the needs of seniors, as the metamodels support only graphical/layout adaptations and do not include multi-modality features such as text-to-speech or speech-to-text. While their evaluation included eight participants over the age of 50 (out of N=23), they did not explore the impact of age-related impairments or user preferences on end-user satisfaction. This omission raises doubts about the applicability of this approach in addressing the diverse and highly personalised accessibility needs of seniors.

In conclusion, almost all existing studies lack the modelling depth necessary to capture the UI accessibility and adaptation needs of seniors at the metamodel layer, with the exception of Bendaly Hlaoui et al. \cite{bendaly2018}. However, even in that case, the comprehensiveness of the models does not necessarily translate to their usability, as evidenced by our experience with developer user studies. For instance, when examining the DSLs and their examples, several key questions arose. To name a few: Are they intuitive for developers? Do they allow developers to embed metadata, such as references to accessibility guidelines and explanations of model functionality? Do they provide flexibility in modelling groups of users via personas rather than individual users? Unfortunately, in most cases, these questions remained unanswered due to the absence of concrete proof-of-concept implementations in the existing studies. Only Yigitbas et al. \cite{yigitbas2020} had such a prototype, but this ties into another major limitation: the lack of evaluation studies capturing insights from software developers on the usability of DSLs and MDE processes. Without such insights, it is difficult to assess how these proposed approaches can be effectively adopted by real-world software developers. 

\section{Contributions} \label{contributions}

\subsection{Contributions to Low-Code Tool Development Community}

\subsubsection{DSL metamodel contributions}

% We made both iterative and novel contributions to the existing metamodels within the MDE adaptive UI subdomain. Context DSL extended an already mature set of approaches and collated the meta classes necessary to map user, platform, and environmental parameters of senior age-related needs with meta classes related to user preferences, assistive technology integration, and input/output devices. Adapt DSL proposed many extensions to previous mature studies that proposed similar adaptation rules via various metadata definition classes, such as comment,s especially providing developers to document which accessibility resource (in our case: Web Content Accessibility Guidelines) influenced a particular adaptation rule. This is also followed by the integration of the UI widget/component structure of an industry-standard cross-platform application framework, such as Flutter can be incorporated into these adaptation rules, so that the developers who define these adaptation within Adapt DSL can use their experience with the integrated UI framework's UI widget/component attributes to define adaptations without a steep learning curve.

\textcolor{black}{We made both iterative and novel contributions to existing metamodels within the MDE adaptive UI domain. Key elements of these contributions are discussed below.}

\vspace{1em}
\noindent\textbf{\textcolor{black}{Context DSL:}}

\begin{itemize}
    \item \textcolor{black}{Context DSL introduces iterative improvements to existing metamodels in the well-established context-of-use modelling approaches within the MDE adaptive UI domain.}
    
    \item \textcolor{black}{We propose that meta-classes related to user preferences, such as text, colour, multimedia, privacy, voiceover, touch, and language, be explicitly integrated into the \textit{UserContext} meta-class in future Context-of-Use DSL implementations.}
    
    \item \textcolor{black}{Additional improvements include extending the \textit{PlatformContext} meta-class with support for assistive technology-related hardware and software features, along with their attributes (e.g., screen readers, voice input, switch control devices, voice controls, screen magnifiers, eye trackers, hearing aids, and wearables), as these technologies are becoming increasingly common in the accessibility landscape.}
    
    \item \textcolor{black}{Furthermore, with the growing adoption of alternative input/output modalities through assistive technologies, we propose extending context-of-use metamodels to map the availability of these modalities to a user or their device by introducing dedicated meta-class extensions to the \textit{PlatformContext} meta-class.}
\end{itemize}

\noindent\textbf{\textcolor{black}{Adapt DSL:}}

\begin{itemize}
    \item \textcolor{black}{The Adapt DSL introduces several significant improvements to existing metamodels dedicated to defining adaptation operations within the MDE adaptive UI domain.}

    \item \textcolor{black}{We propose a richer set of metadata definition classes, including structured \textit{comment} attributes that allow developers to annotate each rule with contextual information such as the underlying problem, proposed solution, associated advantages, and potential trade-offs. Notably, we also support explicit references to accessibility resources that informed a given rule, such as specific criteria from the Web Content Accessibility Guidelines (WCAG).}

    \item \textcolor{black}{We demonstrate that it is both possible and practical to connect abstract app adaptation logic directly with the widget structure of a complex, industry-standard UI framework such as Flutter, at the DSL metamodel level. This approach leverages developers’ existing familiarity with the framework's UI components, their properties, and behaviours, making the DSL easier to learn and more likely to be adopted in real-world projects.}
\end{itemize}

\subsubsection{\toolname~prototype implementation}

\textcolor{black}{Overall, our DSL implementations and the accompanying MDE workflow represent a significant contribution to the MDE-based UI adaptation domain, as they constitute one of the most mature implementations and evaluations currently available within this space. While Yigitbas et al.~\cite{yigitbas2020} present a similarly advanced approach, their work focuses primarily on self-adaptive applications, whereas our contribution is oriented toward design-time adaptation through a developer-focused tool.}

\begin{itemize}
    \item \textcolor{black}{We demonstrate that \textit{Context DSL} can be implemented as a semi-graphical, tree-structured modelling tool to represent the context-of-use parameters for individual seniors or groups of seniors. Compared to existing approaches, we provide empirical evidence that developers find this semi-graphical model both familiar and intuitive due to its resemblance to commonly encountered structures, such as file hierarchies in IDEs and operating systems.}

    \item \textcolor{black}{Our implementation of adaptation rules via \textit{Adapt DSL} highlights the flexibility of the tool in defining both the type and granularity of adaptations. For example, developers can use the DSL to apply a change to a single \textit{Text} widget at the deepest level of the app's abstract syntax tree or modify all \textit{Text} widgets within the UI (or target any level of specificity in between the two extremes). While the current prototype supports adaptations for a dozen or so common widget types, the approach is inherently extensible and can be expanded to include the entire Flutter widget set.}

    \item \textcolor{black}{Another novel contribution lies in the emphasis on metadata during the DSL definition stage, which is preserved and embedded directly into the source code through the MDE process. This feature enhances the explainability and maintainability of the adapted codebase, thereby improving developer experience and supporting the long-term sustainability of low-code projects.}

    \item \textcolor{black}{Finally, \toolname's ability to integrate any Flutter application into the MDE pipeline with minimal integration-related code modifications is a key contribution. For instance, if a developer wishes to increase the text size of specific text widgets, they only need to assign a unique \textit{key} attribute to those widgets. Once an adaptation rule targeting this \textit{key} is defined in the \textit{Adapt DSL} editor, executing the rule automatically applies the corresponding modifications to the source code. This automation is achieved by analysing the Dart abstract syntax tree (AST) to locate the correct insertion points. While reverse-engineering approaches have been explored in related work~\cite{bendaly2018}, our approach is, to our knowledge, the most concrete and the only one applied to a complex UI framework such as Flutter within the MDE-based adaptive UI domain.}
\end{itemize}

\subsection{Contributions for Software Engineering Community}

\begin{itemize}
    \item \textcolor{black}{Unlike much of the related work, where evaluation is often treated as secondary, we adopted a human-centred methodology from the outset and concluded this development cycle with a comprehensive evaluation study. The primary focus of these user studies was on professional software developers, the intended users of \toolname, to understand their perceptions of the tool’s practicality and usefulness. At the same time, we considered the end-user perspective, recognising that developers ultimately build applications for others to use.}
    \begin{itemize}
        \item \textcolor{black}{Developer feedback indicated a strong willingness to adopt \toolname~in real-world software projects. Importantly, criticisms were constructive and focused on suggestions for improvement rather than questioning the tool’s overall value. Based on our analysis, these suggestions are actionable and inform the path toward a future open-source release, contingent upon addressing concerns related to integration throughout different stages of the software development lifecycle.}
    \end{itemize}
    
    \item \textcolor{black}{Based on these findings, we conclude that the software engineering community is likely to find MDE-based UI adaptation approaches such as \toolname~to be both practical and viable for developing more accessible and personalised applications for seniors and other similarly disadvantaged user groups. We believe that such solutions will become increasingly essential in the future, particularly as policy-driven initiatives similar to the European Accessibility Act (EAA)~\cite{eu2019directive882} incentivise software developers to adopt accessible development practices and tools.}
\end{itemize}

\section{Summary} \label{conclusion}

Senior citizens are a crucial part of our society, yet they continue to experience accessibility barriers in apps. The root cause of this issue lies with us, the software practitioners, making it our responsibility to identify solutions that address the age-related accessibility and adaptation needs of seniors. This requires strategies to overcome challenges such as limited time and resources, which are common in real-world software development. In this study, we propose one such strategy: \toolname, a Model-Driven Engineering (MDE) approach that enables developers to automate the generation of adaptive Flutter app instances based on DSL models that define the accessibility-adaptation needs of senior users.

This paper presents the overall architecture, design, and implementation of \toolname. We also describe two user studies conducted with 18 software developers and 22 senior end-users to evaluate the strengths and weaknesses of the current iteration of our tool prototype.

Additionally, we elaborate on the key research contributions of this work. 1) We introduce two novel DSLs that, together, capture age-related accessibility and personalisation needs of senior users. These DSLs are designed to be both comprehensive in their modelling capabilities and accessible to software developers. 2) We propose a novel MDE approach that leverages these DSLs to perform a wide range of source code adaptations in any Flutter application, while integrating seamlessly with existing tools and processes to enhance the developer experience. 3) We conduct a thorough evaluation involving both software developers and senior end-users, providing real-world insights into the design and development of adaptive applications. This evaluation demonstrates the viability and practicality of the approach as a form of developer support tool within the software industry.

% Based on our evaluation, we also present a set of future work to extend our contributions. Namely, iteratively improving \toolname~based on key improvement suggestions received during the evaluation, using this next iterations to allow participants to hands-on experiment with the tool rather than observing the tool via video demonstrations, extending our existing design-time app adaptation workflow to support user-driven multi-modal run-time adaptations, and extending the tool to support serving contextual privacy information to end-users.

\textcolor{black}{Finally, we outline several directions for future work to extend our contributions. These include iteratively improving \toolname~based on key suggestions received during the evaluation; enabling hands-on experimentation in future studies by offering an installable prototype instead of relying solely on video demonstrations; expanding the existing design-time adaptation workflow to support user-driven, multimodal run-time adaptations; and enhancing the tool to provide contextual privacy information to end-users.}

\backmatter

% \bmhead{Supplementary information}

% If your article has accompanying supplementary file/s please state so here. 

% Authors reporting data from electrophoretic gels and blots should supply the full unprocessed scans for key as part of their Supplementary information. This may be requested by the editorial team/s if it is missing.

% Please refer to Journal-level guidance for any specific requirements.

\bmhead{Acknowledgements}

Authors are supported by Australian Research Council (ARC) Laureate Fellowship FL190100035. Our sincere gratitude goes to the participants who took part in the user study.

\section*{Author Contribution}

Shavindra Wickramathilaka: Conceptualisation, Software Development, User Study Conduction, Writing - Original Draft, Review \& Editing. John Grundy: Conceptualisation, Supervision, Writing - Review \& Editing. Kashumi Madampe: Conceptualisation, Supervision, User Study Conduction, Writing - Review \& Editing. Omar Haggag: Conceptualisation, Supervision, Writing - Review \& Editing.

% \begin{appendices}

% \section{Section title of first appendix}\label{secA1}

% An appendix contains supplementary information that is not an essential part of the text itself but which may be helpful in providing a more comprehensive understanding of the research problem or it is information that is too cumbersome to be included in the body of the paper.




% \end{appendices}

\bibliography{bibliography}% common bib file

\end{document}